\input harvmac
\input epsf

%
\let\includefigures=\iftrue
%
%
%
\newfam\black
\input rotate
\input epsf
\noblackbox
%
%
\includefigures
\message{If you do not have epsf.tex (to include figures),}
\message{change the option at the top of the tex file.}
\def\figin{\epsfcheck\figin}\def\figins{\epsfcheck\figins}
\def\epsfcheck{\ifx\epsfbox\UnDeFiNeD
\message{(NO epsf.tex, FIGURES WILL BE IGNORED)}
\gdef\figin##1{\vskip2in}\gdef\figins##1{\hskip.5in}
\else\message{(FIGURES WILL BE INCLUDED)}%
\gdef\figin##1{##1}\gdef\figins##1{##1}\fi}
\def\DefWarn#1{}
\def\N{{\cal N}}
\def\figinsert{\goodbreak\midinsert}
\def\ifig#1#2#3{\DefWarn#1\xdef#1{fig.~\the\figno}
\writedef{#1\leftbracket fig.\noexpand~\the\figno}%
\figinsert\figin{\centerline{#3}}\medskip\centerline{\vbox{\baselineskip12pt
\advance\hsize by -1truein\noindent\footnotefont{\bf
Fig.~\the\figno:} #2}}
\bigskip\endinsert\global\advance\figno by1}
\else
\def\ifig#1#2#3{\xdef#1{fig.~\the\figno}
\writedef{#1\leftbracket fig.\noexpand~\the\figno}%
\global\advance\figno by1} \fi

\def\tilde{\widetilde}

\def\yboxit#1#2{\vbox{\hrule height #1 \hbox{\vrule width #1
\vbox{#2}\vrule width #1 }\hrule height #1 }}
\def\fillbox#1{\hbox to #1{\vbox to #1{\vfil}\hfil}}
\def\ybox{{\lower 1.3pt \yboxit{0.4pt}{\fillbox{8pt}}\hskip-0.2pt}}

\def\rightarrowbox#1#2{
  \setbox1=\hbox{\kern#1{${ #2}$}\kern#1}
  \,\vbox{\offinterlineskip\hbox to\wd1{\hfil\copy1\hfil}
    \kern 3pt\hbox to\wd1{\rightarrowfill}}}

\def\half{{1\over 2}}
\def\Tr{{{\rm Tr~ }}}

\def\Im{{\rm Im\hskip0.1em}}

\def\vev#1{\langle{#1}\rangle}

\def\tilde{\widetilde}

\def\II{\relax{I\kern-.10em I}}

\def\bar{\overline}

\def\IZ{\relax\ifmmode\mathchoice
{\hbox{\cmss Z\kern-.4em Z}}{\hbox{\cmss Z\kern-.4em Z}}
{\lower.9pt\hbox{\cmsss Z\kern-.4em Z}} {\lower1.2pt\hbox{\cmsss
Z\kern-.4em Z}}\else{\cmss Z\kern-.4em Z}\fi}
\def\IB{\relax{\rm I\kern-.18em B}}
\def\IC{{\relax\hbox{$\inbar\kern-.3em{\rm C}$}}}
\def\ID{\relax{\rm I\kern-.18em D}}
\def\IE{\relax{\rm I\kern-.18em E}}
\def\IF{\relax{\rm I\kern-.18em F}}
\def\IG{\relax\hbox{$\inbar\kern-.3em{\rm G}$}}
\def\IGa{\relax\hbox{${\rm I}\kern-.18em\Gamma$}}
\def\IH{\relax{\rm I\kern-.18em H}}
\def\II{\relax{\rm I\kern-.18em I}}
\def\IK{\relax{\rm I\kern-.18em K}}
\def\IN{\relax{\rm I\kern-.18em N}}
\def\IP{\relax{\rm I\kern-.18em P}}

%
\def\inbar{\,\vrule height1.5ex width.4pt depth0pt}

\font\cmss=cmss10 \font\cmsss=cmss10 at 7pt
\def\IR{\relax{\rm I\kern-.18em R}}

\def\lp10{l_P^{10}}
\def\lp11{l_P^{11}}
\def\R11{R_{11}}

\newbox\tmpbox\setbox\tmpbox\hbox{\abstractfont
}
 \Title{\vbox{\baselineskip12pt\hbox to\wd\tmpbox{\hss
 hep-th/0410077} }}
 {\vbox{\centerline{Holomorphic Anomaly Of Unitarity Cuts}
 \bigskip
 \centerline{And One-Loop Gauge Theory Amplitudes}
 }}
\smallskip
\centerline{Freddy Cachazo}
\smallskip
\bigskip
\centerline{\it School of Natural Sciences, Institute for Advanced
Study, Princeton NJ 08540 USA}
\bigskip
\vskip 1cm \noindent

\input amssym.tex

We show how the holomorphic anomaly found in hep-th/0409245 can be
used to efficiently compute certain classes of unitarity cuts of
one-loop $\N=4$ amplitudes of gluons. These classes include all
cuts of n-gluon one-loop MHV amplitudes and of n-gluon next-to-MHV
amplitudes with helicities $(1^+,2^+,3^+,4^-,\ldots , n^-)$. As an
application of this method, we present the explicit computation of
the $(1,2,3)$-cut of the n-gluon one-loop $\N=4$ leading-color
amplitude $A_{n;1}(1^+,2^+,3^+,4^-,\ldots , n^-)$. The answer is
given in terms of scalar box functions and provides information
about the corresponding amplitudes. A possible way to generalize
this method to all kinds of unitarity cuts is also discussed.

\Date{October 2004}

\lref\BernZX{ Z.~Bern, L.~J.~Dixon, D.~C.~Dunbar and
D.~A.~Kosower, ``One Loop N Point Gauge Theory Amplitudes,
Unitarity And Collinear Limits,'' Nucl.\ Phys.\ B {\bf 425}, 217
(1994), hep-ph/9403226.
}

\lref\BernCG{ Z.~Bern, L.~J.~Dixon, D.~C.~Dunbar and
D.~A.~Kosower, ``Fusing Gauge Theory Tree Amplitudes into Loop
Amplitudes,'' Nucl.\ Phys.\ B {\bf 435}, 59 (1995),
hep-ph/9409265.
}

\lref\WittenNN{ E.~Witten, ``Perturbative Gauge Theory as a String
Theory in Twistor Space,'' hep-th/0312171.
}

\lref\CachazoKJ{ F.~Cachazo, P.~Svrcek and E.~Witten, ``MHV
Vertices and Tree Amplitudes in Gauge Theory,'' hep-th/0403047.
}

\lref\berkwitten{N. Berkovits and E. Witten,  ``Conformal
Supergravity In Twistor-String Theory,'' hep-th/0406051.}

\lref\penrose{R. Penrose, ``Twistor Algebra,'' J. Math. Phys. {\bf
8} (1967) 345.}

\lref\berends{F. A. Berends, W. T. Giele and H. Kuijf, ``On
Relations Between Multi-Gluon And Multi-Graviton Scattering,"
Phys. Lett {\bf B211} (1988) 91.}

\lref\berendsgluon{F. A. Gerends, W. T. Giele and H. Kuijf,
``Exact and Approximate Expressions for Multigluon Scattering,"
Nucl. Phys. {\bf B333} (1990) 120.}

\lref\bernplusa{Z. Bern, L. Dixon and D. A. Kosower, ``New QCD
Results From String Theory,'' in {\it Strings '93}, ed. M. B.
Halpern et. al. (World-Scientific, 1995), hep-th/9311026.}

\lref\bernplusb{Z. Bern, G. Chalmers, L. J. Dixon and D. A.
Kosower, ``One Loop $N$ Gluon Amplitudes with Maximal Helicity
Violation via Collinear Limits," Phys. Rev. Lett. {\bf 72} (1994)
2134.}

\lref\bernfive{Z. Bern, L. J. Dixon and D. A. Kosower, ``One Loop
Corrections to Five Gluon Amplitudes," Phys. Rev. Lett {\bf 70}
(1993) 2677.}

\lref\bernfourqcd{Z.Bern and  D. A. Kosower, "The Computation of
Loop Amplitudes in Gauge Theories," Nucl. Phys.  {\bf B379,}
(1992) 451.}

\lref\cremmerlag{E. Cremmer and B. Julia, ``The $N=8$ Supergravity
Theory. I. The Lagrangian," Phys. Lett.  {\bf B80} (1980) 48.}

\lref\cremmerso{E. Cremmer and B. Julia, ``The $SO(8)$
Supergravity," Nucl. Phys.  {\bf B159} (1979) 141.}

\lref\dewitt{B. DeWitt, "Quantum Theory of Gravity, III:
Applications of Covariant Theory," Phys. Rev. {\bf 162} (1967)
1239.}

\lref\dunbarn{D. C. Dunbar and P. S. Norridge, "Calculation of
Graviton Scattering Amplitudes Using String Based Methods," Nucl.
Phys. B {\bf 433,} 181 (1995), hep-th/9408014.}

\lref\ellissexton{R. K. Ellis and J. C. Sexton, "QCD Radiative
corrections to parton-parton scattering," Nucl. Phys.  {\bf B269}
(1986) 445.}

\lref\gravityloops{Z. Bern, L. Dixon, M. Perelstein, and J. S.
Rozowsky, ``Multi-Leg One-Loop Gravity Amplitudes from Gauge
Theory,"  hep-th/9811140.}

\lref\kunsztqcd{Z. Kunszt, A. Singer and Z. Tr\'{o}cs\'{a}nyi,
``One-loop Helicity Amplitudes For All $2\rightarrow2$ Processes
in QCD and ${\cal N}=1$ Supersymmetric Yang-Mills Theory,'' Nucl.
Phys.  {\bf B411} (1994) 397, hep-th/9305239.}

\lref\mahlona{G. Mahlon, ``One Loop Multi-photon Helicity
Amplitudes,'' Phys. Rev.  {\bf D49} (1994) 2197, hep-th/9311213.}

\lref\mahlonb{G. Mahlon, ``Multi-gluon Helicity Amplitudes
Involving a Quark Loop,''  Phys. Rev.  {\bf D49} (1994) 4438,
hep-th/9312276.}

\lref\klt{H. Kawai, D. C. Lewellen and S.-H. H. Tye, ``A Relation
Between Tree Amplitudes of Closed and Open Strings," Nucl. Phys.
{B269} (1986) 1.}

\lref\pppmgr{Z. Bern, D. C. Dunbar and T. Shimada, ``String Based
Methods In Perturbative Gravity," Phys. Lett.  {\bf B312} (1993)
277, hep-th/9307001.}

\lref\GiombiIX{ S.~Giombi, R.~Ricci, D.~Robles-Llana and
D.~Trancanelli, ``A Note on Twistor Gravity Amplitudes,''
hep-th/0405086.
}

\lref\WuFB{ J.~B.~Wu and C.~J.~Zhu, ``MHV Vertices and Scattering
Amplitudes in Gauge Theory,'' hep-th/0406085.
}

\lref\Feynman{R.P. Feynman, Acta Phys. Pol. 24 (1963) 697, and in
{\it Magic Without Magic}, ed. J. R. Klauder (Freeman, New York,
1972), p. 355.}

\lref\Peskin{M.E. Peskin and D.V. Schroeder, {\it An Introduction
to Quantum Field Theory} (Addison-Wesley Pub. Co., 1995).}

\lref\parke{S. Parke and T. Taylor, ``An Amplitude For $N$ Gluon
Scattering,'' Phys. Rev. Lett. {\bf 56} (1986) 2459; F. A. Berends
and W. T. Giele, ``Recursive Calculations For Processes With $N$
Gluons,'' Nucl. Phys. {\bf B306} (1988) 759. }

\lref\BrandhuberYW{ A.~Brandhuber, B.~Spence and G.~Travaglini,
``One-Loop Gauge Theory Amplitudes In N = 4 Super Yang-Mills From
MHV Vertices,'' hep-th/0407214.
}

\lref\CachazoZB{ F.~Cachazo, P.~Svrcek and E.~Witten, ``Twistor
space structure of one-loop amplitudes in gauge theory,''
hep-th/0406177.
}

\lref\passarino{ L.M. Brown and R.P. Feynman, Phys. Rev. 85:231
(1952); G. Passarino and M. Veltman, Nucl. Phys. B160:151 (1979);
G. 't Hooft and M. Veltman, Nucl. Phys. B153:365 (1979);R.G.
Stuart, Comp. Phys. Comm. 48:367 (1988); R.G. Stuart and A.
Gongora, Comp. Phys. Comm. 56:337 (1990).}

\lref\neerven{ W. van Neerven and J.A.M. Vermaseren, Phys. Lett.
137B:241 (1984)}

\lref\melrose{ D.B. Melrose, Il Nuovo Cimento 40A:181 (1965); G.J.
van Oldenborgh and J.A.M. Vermaseren, Z. Phys. C46:425 (1990);
G.J. van Oldenborgh, PhD Thesis, University of Amsterdam (1990);
A. Aeppli, PhD thesis, University of Zurich (1992).}

\lref\bernTasi{Z. Bern, hep-ph/9304249, in {\it Proceedings of
Theoretical Advanced Study Institute in High Energy Physics (TASI
92)}, eds. J. Harvey and J. Polchinski (World Scientific, 1993). }

\lref\morgan{ Z.~Bern and A.~G.~Morgan, ``Supersymmetry relations
between contributions to one loop gauge boson amplitudes,'' Phys.\
Rev.\ D {\bf 49}, 6155 (1994), hep-ph/9312218.
}

\lref\RoiSpV{R.~Roiban, M.~Spradlin and A.~Volovich, ``A Googly
Amplitude From The B-Model In Twistor Space,'' JHEP {\bf 0404},
012 (2004) hep-th/0402016; R.~Roiban and A.~Volovich, ``All Googly
Amplitudes From The $B$-Model In Twistor Space,'' hep-th/0402121;
R.~Roiban, M.~Spradlin and A.~Volovich, ``On The Tree-Level
S-Matrix Of Yang-Mills Theory,'' Phys.\ Rev.\ D {\bf 70}, 026009
(2004) hep-th/0403190.}

\lref\CachazoBY{ F.~Cachazo, P.~Svrcek and E.~Witten, ``Gauge
Theory Amplitudes In Twistor Space And Holomorphic Anomaly,''
hep-th/0409245.
}

\lref\DixonWI{ L.~J.~Dixon, ``Calculating Scattering Amplitudes
Efficiently,'' hep-ph/9601359.
}

\lref\BernMQ{ Z.~Bern, L.~J.~Dixon and D.~A.~Kosower, ``One Loop
Corrections To Five Gluon Amplitudes,'' Phys.\ Rev.\ Lett.\  {\bf
70}, 2677 (1993), hep-ph/9302280.
}

\lref\berends{F.A. Berends, R. Kleiss, P. De Causmaecker, R.
Gastmans and T.T. Wu, Phys. Lett. {\bf B103} (1981) 124; P. De
Causmaeker, R. Gastmans, W. Troost and T.T. Wu, Nucl. Phys. {\bf
B206} (1982) 53; R. Kleiss and W. J. Stirling, Nucl. Phys. {\bf
B262} (1985) 235; R. Gastmans and T.T. Wu, {\it The Ubiquitous
Photon: Heliclity Method For QED And QCD} Clarendon Press, 1990.}

\lref\xu{Z. Xu, D.-H. Zhang and L. Chang, Nucl. Phys. {\bf B291}
(1987) 392.}

\lref\gunion{J.F. Gunion and Z. Kunszt, Phys. Lett. {\bf 161B}
(1985) 333.}

\lref\GeorgiouBY{ G.~Georgiou, E.~W.~N.~Glover and V.~V.~Khoze,
``Non-MHV Tree Amplitudes In Gauge Theory,'' JHEP {\bf 0407}, 048
(2004), hep-th/0407027.
}

\lref\WuJX{ J.~B.~Wu and C.~J.~Zhu, ``MHV Vertices And Fermionic
Scattering Amplitudes In Gauge Theory With Quarks And Gluinos,''
hep-th/0406146.
}

\lref\WuFB{ J.~B.~Wu and C.~J.~Zhu, ``MHV Vertices And Scattering
Amplitudes In Gauge Theory,'' JHEP {\bf 0407}, 032 (2004),
hep-th/0406085.
}

\lref\GeorgiouWU{ G.~Georgiou and V.~V.~Khoze, ``Tree Amplitudes
In Gauge Theory As Scalar MHV Diagrams,'' JHEP {\bf 0405}, 070
(2004), hep-th/0404072.
}

\lref\Nair{V. Nair, ``A Current Algebra For Some Gauge Theory
Amplitudes," Phys. Lett. {\bf B78} (1978) 464. }

\lref\BernAD{ Z.~Bern, ``String Based Perturbative Methods For
Gauge Theories,'' hep-ph/9304249.
}

\lref\BernKR{ Z.~Bern, L.~J.~Dixon and D.~A.~Kosower,
``Dimensionally Regulated Pentagon Integrals,'' Nucl.\ Phys.\ B
{\bf 412}, 751 (1994), hep-ph/9306240.
}

\newsec{Introduction}

The perturbative analysis of gauge theories has been a very
important tool to compare theories and experiments. A great among
of effort has been put on the calculation of scattering amplitudes
in QCD, where the perturbative analysis is useful in the
high-energy regime. Although QCD is well tested in this regime,
calculating these amplitudes is important in order to subtract the
QCD background from possible new physics at colliders.

At tree-level and one-loop level, new techniques have made
possible calculations that are practically impossible by standard
textbook approaches (for a review see for example \DixonWI).

At one-loop level, one such technique is based on the
``supersymmetric" decomposition of QCD amplitudes of gluons.
Namely, a one-loop amplitude in $\N=4$ super Yang-Mills contains
the QCD amplitude plus contributions from fermions and scalars
running in the loop. Combining all fermions with some scalars,
$\N=1$ chiral multiplets can be formed. This leaves only the
contribution of scalars running in the loop. Therefore, once the
supersymmetric amplitudes are known, the complicated QCD
calculation is reduced to that of a scalar running in the loop,
which is much simpler. Thus, calculations of supersymmetric
amplitudes of gluons is also a subject of phenomenological
interest.

One-loop amplitudes of gluons in $\N=4$ gauge theory satisfy three
remarkable properties. The first one is that all integrals that
can appear in a direct Feynman graph calculation can be reduced to
a set of known integrals in dimensional regularization
\refs{\passarino, \neerven, \melrose}. These are known as scalar
box integrals \refs{\bernTasi, \morgan}.

The second property comes from the study of the analytic structure
of scalar box integrals. These integrals are multi-valued
functions, i.e., they have branch cuts in the space of kinematical
invariants. Moreover, there is no linear combination of these
functions, with rational coefficients in the kinematical
invariants, which is single-valued, i.e., a rational function
\refs{\BernZX, \BernCG}.

This gives the second property: All $\N=4$ amplitudes can be
determined completely once their branch cuts and monodromies are
known. Amplitudes with this property are said to be
four-dimensional cut-constructible \refs{\BernZX, \BernCG}.

The third property, also shared by tree-level amplitudes and quite
possibly by higher loops, is that once the amplitudes are
transformed to twistor space, they turn out to be localized on
simple algebraic sets \refs{\WittenNN, \RoiSpV, \CachazoKJ,
\CachazoZB, \CachazoBY}. The algebraic sets can be described as
just unions of ``lines", i.e., $\Bbb{CP}^1$'s linearly embedded in
twistor space, $\Bbb{CP}^3$. At tree-level, all amplitudes were
constructed from a string theory with twistor space as its target
space\foot{There is growing evidence that higher-loop amplitudes
might as well be computed by some sort of string theory in twistor
space.}. This led to a new prescription for calculating {\it all}
tree amplitudes in terms of maximal helicity violating or MHV
amplitudes continued off-shell and connected by Feynman
propagators \CachazoKJ; these are called MHV diagrams.

At one-loop, the simplest twistor picture, originally proposed in
\WittenNN, implies that MHV amplitudes should be localized on two
lines in twistor space \WittenNN. Each line supports a tree-level
MHV amplitude. A straightforward generalization of the tree-level
construction of \CachazoKJ\ suggests that the two lines or
off-shell MHV amplitudes should be connected by two Feynman
propagators to make up the loop amplitude. This idea was
explicitly carried out in \BrandhuberYW\ and found to correctly
reproduce the known n-gluon one-loop MHV amplitudes, first
computed in \BernZX.

On the other hand, the twistor space support of one-loop MHV
amplitudes was studied in \CachazoZB\ by considering the
differential equations they obey. The result involved
configurations were all gluons except one were localized on two
lines. This puzzle was resolved in \CachazoBY; a holomorphic
anomaly affects the result of the differential operators acting on
one-loop amplitudes. The way this anomaly works was most
transparent on the unitarity cuts of one-loop amplitudes
\CachazoBY.

It is the aim of this paper to exploit the holomorphic anomaly to
compute certain unitarity cuts of one-loop $\N=4$ amplitudes. The
importance of this is that once the cuts are know explicitly so is
the amplitude due to the cut-constructibility property.

The basic fact we observe is that any differential operator
designed to annihilate the unitarity cut but that fails to do so
due to the holomorphic anomaly, it is guaranteed to produce a
rational function.

On the other hand, any unitarity cut can be written as the
imaginary part of the amplitude in some suitable kinematical
regime. This implies that it is given by the imaginary part of
some combination of scalar box integrals with rational
coefficients.

We then show that any differential operator that produces a
rational function on the cut via the holomorphic anomaly
annihilates the coefficients of the scalar box integrals in the
amplitude\foot{To be precise, this statement is true only for
scalar box {\it functions}, which are scalar box integrals nicely
normalized.}. This implies that the operator only acts on the
monodromies of the scalar box integrals. These are just logarithms
of rational functions\foot{The four-mass scalar box integral is an
exception to this and it is treated separately.}. Therefore, the
result of applying the operator to the imaginary part is a
rational function with some unknown coefficients.

We give a simple way of comparing the two formulas and therefore
of extracting the unknown coefficients unambiguously. This is
essentially a generalization of the proof of cut-constructibility
of the amplitudes.

Using this method we find the explicit form of the $(1,2,3)$ cut
of the one-loop leading-color partial amplitude
$A_{n:1}(1^+,2^+,3^+,4^-,\ldots ,n^-)$ in terms of only four
scalar box functions.

This paper is organized as follows. In section 2, we briefly
review one-loop amplitudes in $\N=4$ gauge theories and their
cuts. Emphasis is made in the way they can be written in terms of
scalar box functions. In section 2.1, the holomorphic anomaly is
explained in the context of the unitarity cuts. Collinear
operators are introduced and it is explained how they localize the
cut integral to give a rational function. In section 2.2, we give
the recipe for extracting information from this rational function
by comparing it to the action of the collinear operator on the
scalar box functions. The most general classes of cuts in which
this method is directly applicable is also given. These involve
all possible cuts in one-loop MHV amplitudes and in one-loop
next-to-MHV amplitudes with helicities $(1^+,2^+,3^+,4^-,\ldots
,n^-)$.

In section 3, the computation of the $(1,2,3)$ cut of
$A_{n:1}(1^+,2^+,3^+,4^-,\ldots ,n^-)$ is presented following the
general method explained in section 2. In section 4, we make some
consistency checks on the result of section 3. In section 5, we
discuss our results and speculate on the way this can be
generalized to all cuts and thus provide a way of computing {\it
all} one-loop amplitudes. In appendix A, a detailed description of
scalar box functions and their imaginary parts is given. Finally,
appendix B contains some technical details of the computation in
section 3.

\newsec{General Framework}

One-loop amplitudes of gluons in $\N=4$ $U(N)$ gauge theories
depend on the momentum $(p)$, helicity $(h)$, and color index
$(a)$ of each of the external gluons. Consider the n-gluon
amplitude $A_n(\{ p_i,h_i,a_i \})$. It is very useful to separate
the color structure explicitly as follows
\eqn\expu{ A_n(\{ p_i,h_i,a_i \}) =
\sum_{c=1}^{[n/2]+1}\sum_{\sigma} Gr_{n;c}(\sigma
)A_{n;c}(\sigma)}
where
\eqn\human{ Gr_{n;c}(\sigma ) = \Tr (T^{\sigma(a_1)}\ldots
T^{\sigma(a_{c-1})}) \Tr (T^{\sigma(a_c)}\ldots T^{\sigma(a_n)}) }
and $\sigma$ is the set of possible permutations of $n$-gluons mod
out by the symmetries of $Gr_{n;c}(\sigma)$.

This decomposition is useful because the partial amplitudes
$A_{n;c}(\sigma)$ do not have color structure and the Feynman
graphs used for their computation are color-ordered. (For a nice
review see for example \DixonWI.)

Here we study only the leading-color partial amplitudes\foot{It
turns out that subleading-color amplitudes are determined in terms
of the leading-color amplitudes\BernZX.} $A_{n;1}$ for which
$Gr_{n:1}(1) = N \Tr (T^{a_1}\ldots T^{a_{n}})$. In what follows,
we refer to $A_{n;1}$ simply as the $n$-gluon one-loop amplitude.

All n-gluon one-loop MHV amplitudes are known explicitly \BernZX.
These are amplitudes where two gluons have negative (positive)
helicity and $n-2$ have positive (negative) helicity; we refer to
these as mostly plus (mostly minus) MHV amplitudes. Out of the
non-MHV amplitudes, only the simplest one has been computed
\BernCG, i.e., the six-gluon amplitude with three plus and three
minus helicity gluons.

Even though not much is known about general amplitudes, two
important properties are known. One is that all possible Feynman
integrals that can enter in a textbook calculation of these
amplitudes can be expressed in terms of five families of functions
that are explicitly known \BernZX. These are the scalar box
functions (appendix A is devoted to a careful description of these
functions as well as their relation to the scalar box integrals
mentioned in the introduction):
\eqn\fami{\{ F^{1m}_{n:i}, \quad F^{2m~e}_{n:r;i},\quad
F^{2m~h}_{n:r;i}, \quad F^{3m}_{n:r:r';i}, \quad
F^{4m}_{n:r:r':r'';i} \}.}
The range of the indices $\{ r,r',r'',i \}$ depends on the
symmetries of the functions and it is discussed in appendix A.
What is important to us at this point is that all of them are
known very explicitly.

Any n-gluon one-loop amplitude can be written as a linear
combination of the scalar box functions \fami. More explicitly,
\eqn\gene{ \eqalign{& A_{n;1} =  \cr & \sum_{i=1}^n \left( b_i
F^{1m}_{n:i}+ \sum_r c_{r,i} F^{2m~e}_{n:r;i}+ \sum_r d_{r,i}
F^{2m~h}_{n:r;i}+ \sum_{r,r'} g_{\alpha} F^{3m}_{n:r:r';i} +
\sum_{r,r',r''}f_{\beta} F^{4m}_{n:r:r':r'';i} \right)}}
where $\alpha = \{ r,r',i\}$ and $\beta = \{ r,r',r'',i \}$. The
coefficients in this formula naturally have a factor of
$-2ic_{\Gamma}$, where $c_{\Gamma}$ is a constant that shows up
from the dimensional regularization procedure of the amplitude. We
choose to ignore all these overall factors as they can easily be
introduced at the end.

The fact that $A_{n;1}$ can be written as in \gene\ implies that
the task of computing the amplitude is thus reduced to that of
computing the coefficients. These coefficients are rational
functions of the kinematical invariants of the external gluons.
All information about the helicity of the external gluons is
encoded in these coefficients. For example, one-loop MHV
amplitudes have all coefficients either zero or equal to the
corresponding tree-level MHV amplitude \BernZX.

These coefficients are expected to have the simplest form in the
spinor-helicity formalism \refs{\berends,\xu,\gunion} which we
know briefly review. Here we follow the conventions stated in
section 2 of \WittenNN.

In four dimensions, the momentum vector $p$ of a gluon can be
written as a bispinor of the form $p_{a\dot a}
=\lambda_a\tilde\lambda_a$. Spinor inner products are denoted as
$\vev{\lambda, \lambda'} = \epsilon_{ab}\lambda^a\lambda'^b$ and
$[\tilde\lambda, \tilde\lambda'] =\epsilon_{{\dot a}{\dot
b}}\tilde\lambda^{\dot a}\tilde\lambda'^{\dot b}$. In terms of
these, the inner product of the momenta of two gluons, $p$ and $q$
is expressed by $2p\cdot q =
\vev{\lambda_p,\lambda_q}[\tilde\lambda_p, \tilde\lambda_q]$. In
order to avoid cluttering the equations, it is useful to write
$\vev{p~q}$ and $[p~q]$ instead of $\vev{\lambda_p,\lambda_q}$ and
$[\tilde\lambda_p, \tilde\lambda_q]$ respectively.

The main simplification arises because the polarization vector of
a gluon with momentum $p_{a\dot a} =\lambda_a\tilde\lambda_a$ can
be written as: $\epsilon_{a\dot a} = \lambda_a \tilde\nu_{\dot a}/
[\tilde\lambda, \tilde\nu]$ for negative helicity and as
$\epsilon_{a\dot a} = \nu_a\tilde\lambda_{\dot a}/\vev{\nu,
\lambda}$ for positive helicity, where $\nu$ is a fixed reference
spinor. This reference spinor can be wisely chosen to produce
simple formulas. See page 16 of \BernAD\ for an example of such a
simplification in the five-gluon tree-level amplitude.

The final result of this is that the coefficients in \gene\ are
rational functions of spinor products of the external gluons.

\ifig\convi{Representation of the cut integral. Left and right
tree-level amplitudes are on-shell. Internal lines represent the
legs coming from the cut propagators.}
{\epsfxsize=0.50\hsize\epsfbox{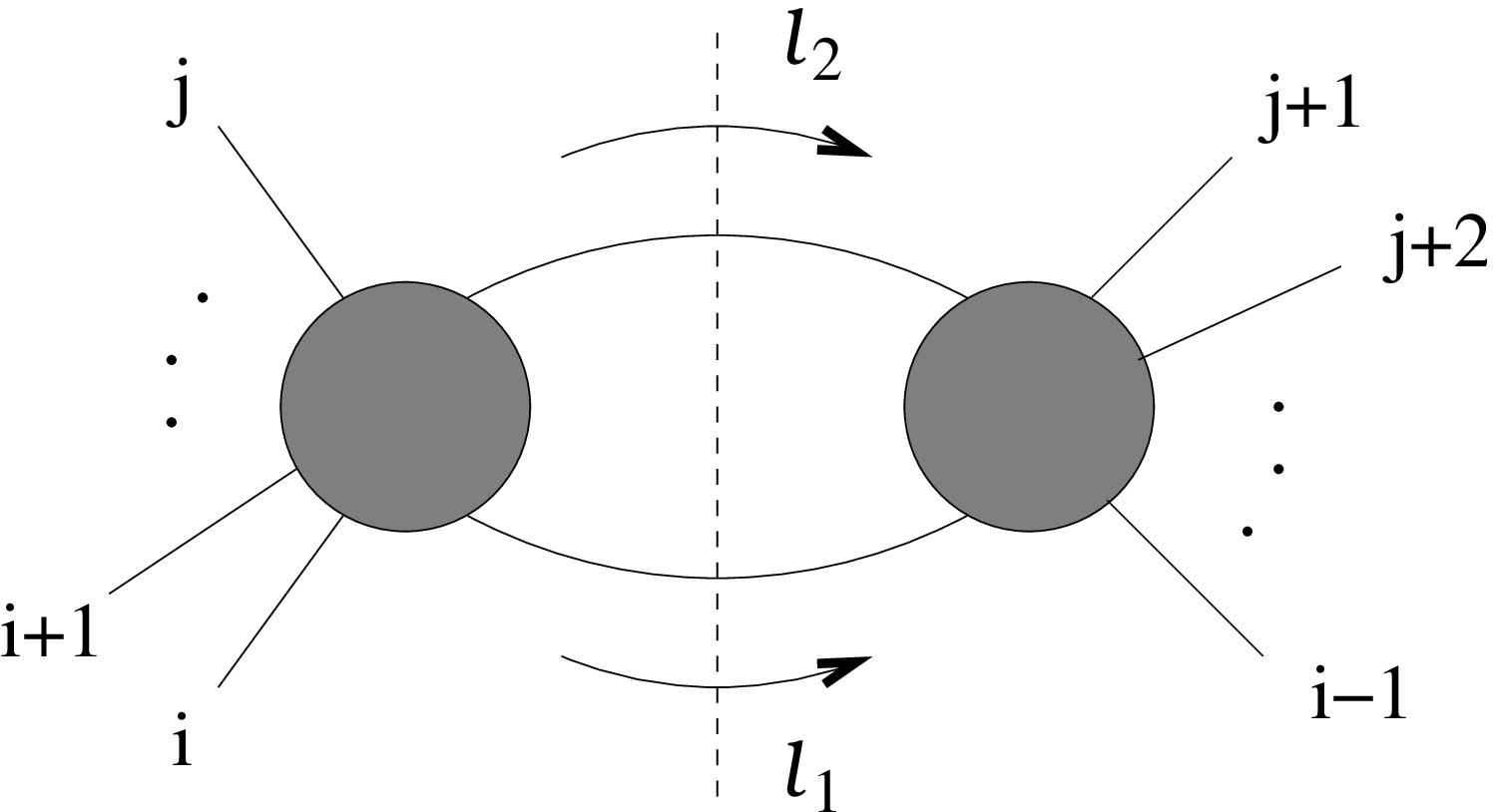}}

The second remarkable property is that these amplitudes are
four-dimensional cut-constructible. This means that their
four-dimensional branch cuts and the corresponding monodromies
determine them uniquely. This is why computing the unitarity cuts
is a way of finding the amplitudes.

Let us turn to the computations of the cuts. Consider for example,
the cut in the $(i,i+1,\ldots, j-1,j)$-channel. This is given by
the ``cut" integral
\eqn\cutIn{ \eqalign{ & C_{i,i+1,\ldots ,j-1,j} = \cr &   \int
d\mu A^{\rm tree}((-\ell_1),i,i+1,\ldots ,j-1,j,(-\ell_2))A^{\rm
tree}(\ell_2,j+1,j+2,\ldots ,i-2,i-1,\ell_1)}}
where $d\mu$ is the Lorentz invariant phase space measure of two
light-like vectors $(\ell_1, \ell_2)$ constrained by momentum
conservation. We find it useful to define $\ell_1$ and $\ell_2$ as
in \convi. We think about the flow of energy as going from one
side of the cut (dashed line in \convi) to the other. Helicity
assignments are for incoming particles\foot{These conventions are
slightly different from those used in \BernZX\ but, of course, the
results are independent of the choice.}.

These cuts \cutIn\ compute the imaginary part of the amplitude in
some suitable chosen kinematical regime. To describe this regime
more explicitly we have to consider the possible kinematical
invariants that the scalar box functions can depend on. Due to the
color-order, the only invariants involve consecutive sets of
gluons and are usually denoted by $t_i^{[r]} = (p_i+p_{i+1}+ ...
+p_{i+r-1})^2$. In other words, the invariants are characterized
by chains of gluons labelled by the first gluon in the chain $(i)$
and by the length of the chain $(r)$.

The cut $C_{i,i+1,\ldots ,j-1,j}$ computes the imaginary part of
\gene\ in the unphysical kinematical regime where $t_i^{[j-i+1]} =
(p_i+p_{i+1}+\ldots +p_j)^2$ is positive and all other invariants
are negative \BernZX.

It is now clear that computing $C_{i,i+1,\ldots ,j-1,j}$ provides
information about the coefficients in \gene\ via
\eqn\word{C_{i,i+1,\ldots ,j-1,j} = {\rm Im}|_{t_{i}^{[j-i+1]}>0}
A_{n;1}.}

Indeed, this is the way that all known one-loop $\N=4$ amplitudes
have been computed or checked \refs{\BernZX,\BernCG}.

We know turn to the problem of computing $C_{i,i+1,\ldots ,j-1,j}$
using the holomorphic anomaly found in \CachazoBY.

\subsec{Twistor Space Support, Collinear Operators And Holomorphic
Anomaly}

In \WittenNN, a remarkable conjecture about the localization of
$\N=4$ amplitudes in twistor space was made. Amplitudes of $n$
gluons at $L$-loop order with $q$ gluons of negative helicity were
conjectured to be localized on curves of genus $g\leq L$ and
degree $d=q+L-1$ when transformed to twistor space.

This conjecture was further explored in \refs{\RoiSpV, \CachazoKJ,
\CachazoZB, \CachazoBY}. In particular, it was shown in
\CachazoKJ\ that all tree-level amplitudes could be computed from
configurations of unions of $q-1$ lines, i.e. $\Bbb{CP}^1$'s.
Mostly plus MHV one-loop amplitudes were considered in \CachazoZB.
They turn out to be localized on unions of lines after a
holomorphic anomaly in the analysis is taken into account
\CachazoBY. This is in perfect agreement with the original picture
of \WittenNN.

In \WittenNN, a method for testing localization of gluons in
twistor space was proposed. Suppose that gluons $i$,$j$, and $k$
are collinear in twistor space\foot{This collinearity condition
has nothing to do with the usual meaning of the word collinear in
the scattering amplitude literature. There, two gluons are
collinear if the corresponding momenta are proportional.}. In
other words, the twistor transform of the amplitude vanishes
unless these gluons lie on the same $\Bbb{CP}^1$ inside
$\Bbb{CP}^3$. This means that if $Z^I =(\lambda^a,\mu^{\dot a})$,
with $\mu_{\dot a} = -i\del /\del \tilde\lambda^{\dot a}$, are
coordinates in twistor space, then the vector $V_L =
\epsilon_{IJKL}Z^I_iZ^J_jZ^K_k$ vanishes. In Minkowski space, this
is a vector of differential operators; each component is supposed
to annihilate any amplitude in which gluons $i$,$j$, and $k$ are
collinear in twistor space.

Let us choose the dotted components, $L=\dot a$, and construct the
following spinor-valued operator
\eqn\spin{ F_{ijk;\dot a} = \epsilon_{IJK\dot a}Z^I_iZ^J_jZ^K_k .}
More explicitly,
\eqn\expis{ F_{ijk;\dot a} = \vev{i~j}{\del \over \del
\tilde\lambda_k^{\dot a}} + \vev{k~i}{\del \over \del
\tilde\lambda_j^{\dot a}} + \vev{j~k}{\del \over \del
\tilde\lambda_i^{\dot a}}.}
The advantage of this choice if that the operator \expis\ is of
first order while for $L=a$ is of second order.

In the following, it will be convenient to introduce a fixed
arbitrary negative chirality spinor $\eta^{\dot a}$ and consider
\eqn\prox{ [F_{ijk}, \eta] = \epsilon^{\dot a\dot b}\eta_{\dot
a}F_{ijk;\dot b}. }
Note that the brackets in \prox\ are meant to indicate the inner
product of two negative chirality spinors and not the commutator
of operators.

A very important example is when $[F_{ijk}, \eta]$ acts on a
function $G$ whose dependence on the three gluons $i$, $j$, and
$k$ is only through $\{ \lambda_i,\lambda_j,\lambda_k,
(p_i+p_j+p_k) \}$. We want to show that $[F_{ijk}, \eta]G = 0$.
Upon using the chain rule we find that $[F_{ijk}, \eta]G$ is
proportional to the positive chirality spinor
\eqn\collix{\nu = \vev{i~j}\lambda_k + \vev{k~i}\lambda_j
+\vev{j~k}\lambda_i.}
The idea is to show that $\nu$ is zero. This is equivalent to
showing that $\vev{\nu, \chi}$ is zero for any $\chi$. But this is
exactly equal to Schouten's identity
\eqn\schouten{ \vev{i~j}\vev{k~\chi} + \vev{k~i}\vev{j~\chi}
+\vev{j~k}\vev{i~\chi} = 0.}

In this derivation we have assumed that the operator acts
trivially on $\lambda_i,\lambda_j,\lambda_k$. However, as shown in
\CachazoBY, this might not be the case when $G$ is a one-loop
amplitude or its unitarity cut.

In \CachazoBY, it was shown that at one-loop there are situations
in which gluons $i$, $j$, and $k$ are collinear in twistor space
and yet the operator \spin\ does not annihilate the amplitude.
This was most clearly explained by considering not the amplitude
but its unitarity cuts \cutIn.

Consider the cut \cutIn. For simplicity, let us assume that the
particles in the loop are gluons. This restriction is useful
because both tree-level amplitudes in \cutIn\ only involve gluons
and can be expanded in terms of the MHV diagrams of \CachazoKJ.
Moreover, in the main example of this paper discussed in the next
section, only gluons can actually propagate in the loop. The case
when scalars or fermions run in the loop is slightly more
complicated, for it involves generalizations of MHV diagrams
\refs{\GeorgiouWU,\WuFB,\WuJX,\GeorgiouBY}.

As explained in \CachazoBY, the holomorphic anomaly only affects
the action of collinear operators that involve gluons next to the
internal ones (for example, gluons $i-1$, $i$, $j$, and $j+1$ in
\cutIn). The reason is that the anomaly shows up only when the
integrant of \cutIn\ has a pole when the momentum of an internal
gluon becomes proportional that of an external gluon. Due to the
color-order, this only happens for adjacent gluons.

The classes of cuts we are interested in this paper are those for
which any of the special gluons is collinear (in twistor space)
with two more external gluons.

More explicitly, cuts belonging to this class\foot{See section 5
for possible generalizations.} are those for which one of the
tree-level amplitudes in \cutIn\ is a mostly plus MHV amplitude.
Recall that mostly plus tree-level MHV amplitudes are manifestly
localized on lines \WittenNN. Therefore, all gluons participating
in the cut are collinear. Of course, at least three external
gluons should participate in order to satisfy the
criterion\foot{Two-particle cuts are special. They have
singularities and make sense only in some regularization scheme.
However, no scalar box function has cuts {\it only} in
two-particle channels; therefore, studying all other channels must
suffice to determine the whole amplitude.}.

Let the left tree amplitude in \cutIn\ be the mostly plus MHV
amplitude. Assume that gluons $k$ and $m$ have negative helicity,
\eqn\culi{\eqalign{ & C_{i,i+1,\ldots ,j-1,j} = \cr &  \int d\mu\;
A^{\rm tree MHV}_{km}((-\ell_1),i,(i+1),\ldots ,j,(-\ell_2))A^{\rm
tree}(\ell_2,j+1,j+2,\ldots ,i-1,\ell_1)} }
where gluons $j+1$ through $i-1$ can have any helicity. Note that
$k$ and $m$ can be any gluons, including $\ell_1$ and $\ell_2$.

Let us write the left tree amplitude explicitly \parke,
\eqn\treeMHV{A^{\rm tree MHV}_{km}((-\ell_1),i,(i+1),\ldots
,j,(-\ell_2)) = {\vev{k~m}^4\over \vev{\ell_1, i}\vev{i~i+1}\cdot
\vev{j-1~j}\vev{j~\ell-2}}.}
Using this in \cutIn\ we have
\eqn\didi{\eqalign{ & C_{i,i+1,\ldots ,j-1,j} = \cr & \int d\mu
{\vev{k~m}^4\over \vev{i~i+1}\ldots \vev{j-1~j}} {1\over
\vev{\ell_1~i}\vev{j~\ell_2}}A^{\rm tree}(\ell_2,j+1,j+2,\ldots
,i-1,\ell_1) .}  }

Now consider the action of the collinear operator \prox\ for
gluons $i$, $i+1$ and $i+2$, i.e.,
\eqn\kumba{  [F_{i,i+1,i+2},\eta] =
\vev{i~i+1}[\del_{i+2},\eta]+\vev{i+1~i+2}[\del_{i},\eta]+\vev{i+2~i}[\del_{i+1},\eta]
}
where $\del_k = \del/\del \tilde\lambda_k$.

Naively, $[F_{i,i+1,i+2},\eta]C_{i,i+1,\ldots ,j-1,j}=0$, but as
pointed out in \CachazoBY, the presence of the pole $1/
\vev{\ell_1~i}$ makes the action of \kumba\ nontrivial.

The basic idea is that the action of $[F_{i,i+1,i+2},\eta]$ on the
pole produces a delta function via\foot{This formula is not
directly applicable to $[F_{i,i+1,i+2},\eta]C_{i,i+1,\ldots
,j-1,j}$. A Schouten identity \schouten\ has to applied to go from
one to the other. See appendix B for more details.}
\eqn\really{ [d\bar\lambda_{\ell_1}, \del_{\ell_1}]{1\over
\langle\lambda_{\ell_1},\lambda_i\rangle} =
d\bar\lambda_{\ell_1}^{\dot
a}{\partial\over\partial\bar\lambda_{\ell_1}^{\dot a}}{1\over
\langle\lambda_{\ell_1},\lambda_i\rangle}=
2\pi\bar\delta(\langle\lambda_{\ell_1},\lambda_i\rangle),}
where we have introduced a $(0,1)$-form $\bar\delta (z) =-id\bar z
\;\delta^2(z)$.

The main simplification arises because the integral over the
Lorentz invariant phase space is an integral over a two sphere and
the delta function produced by $[F_{i,i+1,i+2},\eta]$ is enough to
localize the integral completely. In other words, evaluating the
action of $[F_{i,i+1,i+2},\eta]$ on $C_{i,i+1,\ldots ,j-1,j}$ does
not involve any actual integration.

The localization produced by the delta function turns out to set
$\ell_1 =t p_i$ and $\ell_2 = P_L - tp_i$ where $P_L = p_i +
\ldots +p_j$ and $t = t_i^{[j-i+1]}/ (2p_i\cdot P_L)$. This is
shown in appendix B.

Therefore, the result of the action of the collinear operator on
the cut is schematically
\eqn\excutre{\eqalign{ & [F_{i,i+1,i+2},\eta]C_{i,i+1,\ldots
,j-1,j} = \cr & {\cal J}\times {\vev{k~m}^4\over \vev{i~i+1}\ldots
\vev{j-1~j}} {1\over \vev{j~\ell_2}}A^{\rm
tree}(\ell_2,j+1,j+2,\ldots ,i-1,\ell_1)}}
with $\ell_1$ and $\ell_2$ set to the values given above and
${\cal J}$ is a Jacobian factor that needs to be computed\foot{We
have included the factor of $\vev{i+1~i+2}$ from the collinear
operator in the definition of the Jacobian ${\cal J}$.}.

This proves that the action of $[F_{i,i+1,i+2},\eta]$ on
$C_{i,i+1,\ldots ,j-1,j}$ is a rational function\foot{Recall that
all tree-level amplitudes are rational functions.}.

Important examples of one-loop amplitudes with {\it all} cuts
satisfying the criterion stated above are MHV amplitudes and the
next-to-MHV amplitudes $A_{n:1}(1^+,2^+,3^+,4^-,\ldots , n^-)$. In
section 3, we compute $C_{123}$ of the latter.

\subsec{Extracting Information From Collinear Operators Acting On
Cuts}

We have shown that the evaluation of certain collinear operators
on cuts of the form \culi\ is very simple. As claimed in the
introduction, $[F_{i,i+1,i+2},\eta]C_{i,i+1,\ldots ,j-1,j}$ is a
rational function.

The question is whether this can be used to obtain the explicit
form of the cut in terms of the imaginary part of scalar box
functions.

The idea is to apply $[F_{i,i+1,i+2},\eta]$ to \word, i.e,
\eqn\wordy{ [F_{i,i+1,i+2},\eta]C_{i,i+1,\ldots ,j-1,j} =
[F_{i,i+1,i+2},\eta]\; {\rm Im}|_{t_{i}^{[j-i+1]}>0} A_{n;1}.}
with $A_{n:1}$ given by \gene. Generically, all coefficients of
\gene\ are unknown.

It turns out that the imaginary part of any scalar box function is
of the form $\ln G$ where $F$ is some rational function of the
momentum invariants. To be more precise, this is the case for all
scalar box functions except for the four-mass function, where $G$
can have the form $F+\sqrt{K}$, with $F$ and $K$ rational
functions (see appendix A). At the end of appendix A, we prove
that this rules out all four-mass box functions in cuts of the
form \culi.

The collinear operator \prox\ is a first order differential
operator. Therefore, it produces two terms for each term in ${\rm
Im}|_{t_{i}^{[j-i+1]}>0} A_{n:1}$. Namely, one term when it acts
on the coefficient of a given scalar box function and one more
when it acts on the logarithm, i.e., the imaginary part of the
scalar box function. When the collinear operator acts on the
logarithms, it produces a rational function\foot{This is where the
four-mass function is different from the others. See appendix A.}.
On the other hand, when it acts on the coefficients, the logarithm
survives.

There are only two ways this can be consistent with the fact that
$[F_{i,i+1,i+2},\eta]C_{i,i+1,\ldots ,j-1,j}$ must be a rational
function. One possibility is that $[F_{i,i+1,i+2},\eta]$ gives
zero when acting on the coefficients. The other possibility is
that terms coming from different box functions conspire to cancel
the logarithms. We now prove that the latter possibility is ruled
out.

The proof goes along the same lines as the proof of the
cut-constructibility of $\N=4$ amplitudes in \BernZX. We take the
scalar box functions and consider the limit when the only positive
kinematical invariant is large. In this limit, the imaginary part
of each of the scalar box function develops a unique function of
the form
\eqn\devel{ {\rm Im}|_{t_i^{[r]}>0}\left(
\ln(-t_i^{[r]})\ln(-t_{i'}^{[r']})\right) = \pi
\ln(-t_{i'}^{[r']}).}
Upon applying a collinear operator to \devel, one gets a pole, $1/
t_{i'}^{[r']}$, that could be called the ``signature" of the
corresponding scalar box function in this channel.

Let us study all scalar box functions one at a time. Recall that
we only consider cuts in more than two-particle channels (see
footnote on page 9).

\item{1.} The three-mass box function, $F^{3m}_{n:r:r';i}$. This
function participates in four cuts:
\eqn\threeM{\eqalign{& {\rm Im}|_{t_i^{[r]}>0}\left(
\ln(-t_{i}^{[r]})\ln(-t_{i+r+r'}^{[n-r-r'-1]})\right) = \pi
\ln(-t_{i+r+r'}^{[n-r-r'-1]}), \cr & {\rm
Im}|_{t_{i+r+r'}^{[n-r-r'-1]}>0}\left(
\ln(-t_i^{[r]})\ln(-t_{i+r+r'}^{[n-r-r'-1]})\right) = \pi
\ln(-t_{i}^{[r]}), \cr & {\rm Im}|_{t_{i-1}^{[r+1]}>0}\left(
\ln(-t_{i-1}^{[r+1]})\ln(-t_{i}^{[r+r']})\right) = \pi
\ln(-t_{i}^{[r+r']}), \cr & {\rm Im}|_{t_i^{[r+r']}>0}\left(
\ln(-t_{i-1}^{[r+1]})\ln(-t_{i}^{[r+r']})\right) = \pi
\ln(-t_{i-1}^{[r+1]}). } }

\item{2.} The two-mass ``easy" function, $F^{2m~e}_{n:r;i}$. This
function also participates in four cuts:
\eqn\twoeasy{\eqalign{& {\rm Im}|_{t_i^{[r]}>0}\left(
\ln(-t_{i}^{[r]})\ln(-t_{i+r+1}^{[n-r-2]})\right) = \pi
\ln(-t_{i+r+1}^{[n-r-2]}), \cr & {\rm
Im}|_{t_{i+r+1}^{[n-r-2]}>0}\left(
\ln(-t_i^{[r]})\ln(-t_{i+r+1}^{[n-r-2]})\right) = \pi
\ln(-t_{i}^{[r]}), \cr & {\rm Im}|_{t_{i-1}^{[r+1]}>0}\left(
\ln(-t_{i-1}^{[r+1]})\ln(-t_{i}^{[r+1]})\right) = \pi
\ln(-t_{i}^{[r+1]}), \cr & {\rm Im}|_{t_i^{[r+1]}>0}\left(
\ln(-t_{i-1}^{[r+1]})\ln(-t_{i}^{[r+1]})\right) = \pi
\ln(-t_{i-1}^{[r+1]}). }   }

\item{3.} The two-mass ``hard" function, $F^{2m~h}_{n:r;i}$. This
function only participates in three cuts:
\eqn\twohard{\eqalign{& {\rm Im}|_{t_i^{[r]}>0}\left(
\ln(-t_{i}^{[r]})\ln(-t_{i-1}^{[r+1]})\right) = \pi
\ln(-t_{i-1}^{[r+1]}), \cr & {\rm Im}|_{t_{i-1}^{[r+1]}>0}\left(
\ln(-t_{i-1}^{[r+1]})\ln(-t_{i-2}^{[2]})\right) = \pi
\ln(-t_{i-2}^{[2]})  \cr & {\rm Im}|_{t_{i+r}^{[n-r-2]}>0}\left(
\ln(-t_{i+r}^{[n-r-2]})\ln(-t_{i-1}^{[r+1]})\right) = \pi
\ln(-t_{i-1}^{[r+1]}). }   }

\item{4.} The one-mass function, $F^{1m}_{n;i}$. This function
only participates in one cut:
\eqn\oneM{ {\rm Im}|_{t_{i-3}^{[3]}>0}\left(
\ln(-t_{i-3}^{[3]})\ln(t_{i-3}^{[2]}t_{i-2}^{[2]})\right) = \pi
\ln(t_{i-3}^{[2]}t_{i-2}^{[2]}). }

By inspection it is clear that the corresponding ``signatures" in
a given channel are indeed unique.

There is one possible problem with this argument. Suppose that one
chooses a collinear operator that does not annihilate a given
scalar box function but it annihilates its ``signature". In such a
case, one has to be more careful and look for other ways to single
out the corresponding box function.

In order to make use of these unique signatures, we have to show
that there exits at least one collinear operator that is
completely safe from the problem mentioned above. Consider any
collinear operator made out of gluons in only the left side of the
cut. Such operator annihilates the kinematical invariant which is
positive and large. It is easy to see that such operator does not
annihilate any of the signatures unless it annihilates the whole
box function.

This concludes our proof that rules out the possibility of a
conspiracy.

Finally, all we have to do is to find out a way of matching the
two rational functions, \wordy\ and \culi, in order to extract the
coefficients. Clearly, the idea is to look for the ``signatures"
that appear in \wordy\ in the action of $[F_{i,i+1,i+2},\eta]$ on
\culi. We illustrate this procedure in the rest of the paper by
computing the $(1,2,3)$ cut of $A_{n:1}(1^+,2^+,3^+,4^-,\ldots
,n^-)$.

\bigskip
\noindent {\it A Subtlety}

As mentioned above, sometimes some of the scalar box functions
participating in the cut $C_{i,i+1,\ldots ,j-1,j}$ are annihilated
by the collinear operator $[F_{i,i+1,i+2},\eta]$. Therefore, no
information can be obtained about the corresponding coefficients
from this operator.

The way to solve this problem is to apply a different operator
that does not annihilate the scalar box functions that the first
operator annihilated. The example we consider in the next section
exhibits this generic behavior.

\newsec{Computation Of The $t_{1}^{[3]}$ Cut Of $A^{\rm one-loop}_{n:1}(1^+,2^+,3^+,4^-,5^-,\ldots , n^-)$}

In this section we present the analysis of the $t_1^{[3]}$ cut of
the n-gluon one-loop $\N=4$ amplitude
$A_{n:1}(1^+,2^+,3^+,4^-,5^-,\ldots , n^-)$. As discussed in the
previous section, the full amplitude is a sum over box functions.
Here we compute the coefficients of all the box functions that
have a branch cut in the $t_1^{[3]}$ channel. In general, there
are $(n-2)(n-3)/2$ scalar box functions with cuts in this channel.

\ifig\lipu{Cut integral in the $t_1^{[3]}$ channel.}
{\epsfxsize=0.45\hsize\epsfbox{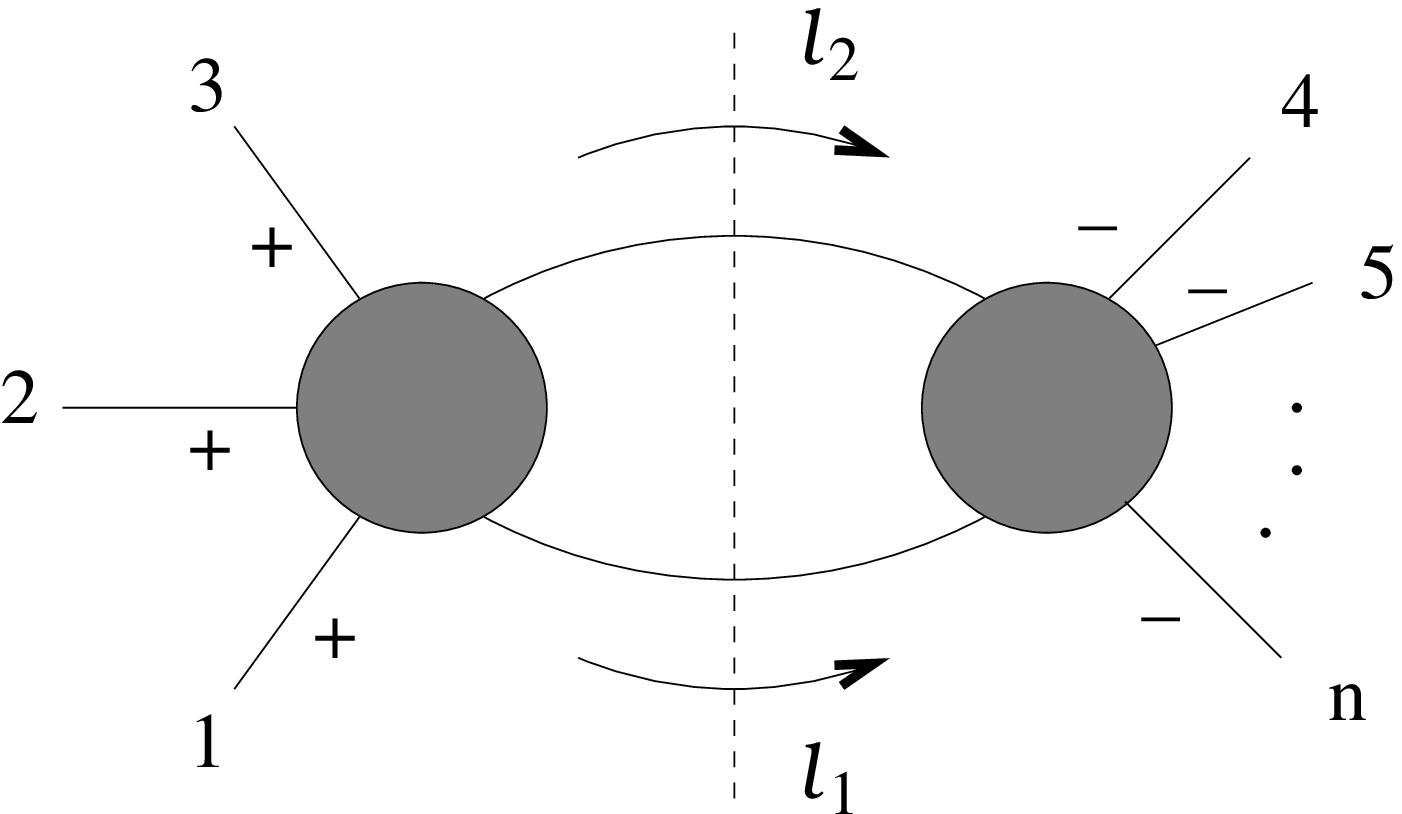}}

The cut is explicitly given by
\eqn\hiho{\eqalign{ C_{123} & =  \int d\mu A^{\rm
tree}((-\ell_1)^-,1^+,2^+,3^+,(-\ell_2)^-)A^{\rm
tree}(\ell_2^+,4^-,5^-,\ldots ,n^-,\ell_1^+) \cr & = -{1\over
\vev{1~2}\vev{2~3}[4~5][5~6]\ldots [n-1 ~ n]}\int d\mu
{\vev{\ell_1~\ell_2}^3[\ell_1~\ell_2]^3\over \vev{\ell_1~
1}\vev{3~\ell_2}[\ell_2~4][n~\ell_1]} \cr & = -{(t_1^{[3]})^3\over
\vev{1~2}\vev{2~3}[4~5][5~6]\ldots [n-1 ~ n]} \int d\mu {1\over
\vev{\ell_1~ 1}\vev{3~\ell_2}[\ell_2~4][n~\ell_1]} ,} }
where $d\mu$ is a measure over the Lorentz invariant phase space
of $(\ell_1,\ell_2)$, which we write down explicitly below. Also
note that in this case, gluons are the only particles that can run
in the loop. Moreover, they can only have the helicities given in
\hiho. The reason is that one of the tree-level amplitudes in
\hiho\ is zero for internal scalars, fermions, or gluons with a
different helicity assignment (see for example \DixonWI).

If the amplitude was known, $C_{123}$ could be computed by taking
the imaginary part of it in the kinematical regime where
$t_1^{[3]}>0$ and all other invariants are negative. More
explicitly, the cut would be given by
\eqn\gola{\eqalign{ C_{123} = {\rm Im}|_{t_1^{[3]}>0} & ~  \left(
b_4 F^{1m}_{n:4} + c_{2,2} F^{2m~e}_{n:2;2} + c_{3,1}
F^{2m~e}_{n:3;1} + c_{2,1} F^{2m~e}_{n:2;1} + d_{2,2}
 F^{2m~h}_{n:2;2} + d_{3,1} F^{2m~h}_{n:3;1} \right. \cr &  +
d_{n-5,6} F^{2m~h}_{n:n-5;6}+d_{n-4,5} F^{2m~h}_{n:n-4;5} +
\sum_{r'=2}^{n-5} g_{2,r',2} F^{3m}_{n:2:r';2} + \sum_{r'=2}^{n-6}
g_{3,r',1} F^{3m}_{n:3:r';1}\cr &  + \sum_{i=7}^{n-1}
g_{n-i-1,3,i} F^{3m}_{n:n-i-1:3;i} + \sum_{r=2}^{n-6}
g_{r,n-r-4,5} F^{3m}_{n:r:n-r-4;5} + \cr & \left. \sum_{r=2}^{n-4}
g_{r,n-r-3,4} F^{3m}_{n:r:n-r-3;4}  +
\sum_{r'=2}^{n-7}\sum_{r"=2}^{n-r'-5} f_{3,r',r'',1}
F^{4m}_{n:3:r':r'';1} \right).}}
We have written only the $(n-2)(n-3)/2$ scalar box functions that
develop an imaginary part.

The imaginary part of each of these functions can be computed
using the formulas given in appendix A.

The main result of this section is the computation of all
$(n-2)(n-3)/2$ coefficients in \gola. The result is strikingly
simple and is given by\foot{Recall that we are omitting all
constants that appear as overall factors, including factors of $i$
and $2\pi$. These can be easily worked out if needed.}
\eqn\resur{ C_{123} = {\cal B}_n ~{\rm Im}|_{t_1^{[3]}>0} \left(
F^{1m}_{n:4} +  F^{2m~e}_{n:3;1} + F^{2m~h}_{n:2;2} +
F^{2m~h}_{n:n-4;5} \right)}
where
\eqn\defB{ {\cal B}_n = {(t_1^{[3]})^3\over
\vev{1~2}\vev{2~3}[4~5][5~6]\ldots [n-1 ~ n] {\langle 1|1+2+3|4
]}{\langle 3|1+2+3|n]} } }
and ${\langle i | 1+2+3 | k] = \vev{i~1}[1~k]+ \vev{i~2}[2~k]+
\vev{i~3}[3~k]}$.

\subsec{Collinear Operators Acting On $C_{123}$}

The explicit form of $C_{123}$ in \hiho\ makes it manifest that
its twistor support has gluons $1,2,3$ on the same ``line" or
$\Bbb{CP}^1$, i.e., they are collinear. This is because mostly
plus tree-level MHV amplitudes are localized on a line \WittenNN.
If there was no holomorphic anomaly, this localization would imply
that the differential operator $[F_{123},\eta]$ annihilates
$C_{123}$. According to the general discussion of section 2, this
is precisely the kind of operators that can give the most
information about the cut.

Let us compute $[F_{123},\eta]C_{123}$ explicitly using the
holomorphic anomaly. The computation starts by applying
$[F_{123},\eta]$ to $C_{123}$ given in \hiho. As expected, we find
that the operator only acts on the poles,
\eqn\pola{ \eqalign{ & [F_{123},\eta] C_{123} =
{(t_1^{[3]})^3\over \vev{1~2}\vev{2~3}[4~5][5~6]\ldots [n-1 ~ n]}
\times \cr & \int d\mu \left[ {1\over \vev{3~
\ell_2}[\ell_2~4][n~\ell_1]}\vev{2~3}[\del_1,\eta]\left( {1\over
\vev{\ell_1~1}}\right) + {1\over \vev{\ell_1
~1}[\ell_2~4][n~\ell_1]}\vev{1~2}[\del_3,\eta]\left( {1\over
\vev{3~\ell_2}}\right) \right].} }
To see this note that $[F_{123},\eta]t_1^{[3]} = 0$ due to
\schouten. Recall that $\del_k = \del/\del\tilde\lambda_k$.

Let us concentrate on the computation of the first term leaving
out the overall factor containing $t_1^{[3]}$. In order to avoid
cluttering the formulas, it is convenient to introduce
\eqn\intros{  P = p_1 + p_2 + p_3.}

We have to write the Lorentz invariant measure explicitly
\eqn\kawa{ \eqalign{ & \int d\mu  {1\over \vev{3~
\ell_2}[\ell_2~4][n~\ell_1]}\vev{2~3}[\del_1,\eta]\left( {1\over
\vev{\ell_1~1}}\right) = \cr & \int d^4\ell_1
\delta^{(+)}(\ell_1^2)\int d^4\ell_2 \delta^{(+)}(\ell_2^2)
\delta^{(4)}(\ell_1 + \ell_2-P) {1\over \vev{3~
\ell_2}[\ell_2~4][n~\ell_1]}\vev{2~3}[\del_1,\eta]\left( {1\over
\vev{\ell_1~1}}\right).} }
As mentioned above, counting the two delta functions produced by
the holomorphic anomaly, i.e., by the action of $[\del_1,\eta]$ on
the pole, we have a total of eight delta functions. These are
enough to localize the integral \kawa\ completely, i.e., no actual
integration has to be performed.

In appendix B, we provide a detailed computation of \kawa. Let us
summarize the results of the localization: $\ell_1 = t p_1$,
$\ell_2 = P-t p_1 $, with $t= P^2/(2p_1\cdot P)$. The evaluation
in appendix B shows that \kawa\ is
\eqn\ffool{\int d\mu  {1\over \vev{3~
\ell_2}[\ell_2~4][n~\ell_1]}\vev{2~3}[\del_1,\eta]\left( {1\over
\vev{\ell_1~1}}\right) =  {[1~\eta]\over {\langle 1
|P|4]}[n~1][1~2]}. }
This is essentially the general form given in \excutre\ with the
jacobian explicitly computed.

The computation of the second term in \pola\ is completely
similar. Combining the two results, we obtain the final form for
the action of the collinear operator on the cut
\eqn\grow{ [F_{123},\eta] C_{123} = {(t_1^{[3]})^3\over
\vev{1~2}\vev{2~3}[4~5][5~6]\ldots [n-1 ~ n]}  \left(
{[1~\eta]\over {\langle 1 |P|4]}[n~1][1~2]} + {[3~\eta]\over
{\langle 3 |P|n]}[2~3][3~4]} \right) .}

\bigskip
\noindent{\bf ``Simple Fraction" Expansion}

Although our formula \grow\ for $[F_{123},\eta] C_{123}$ is quite
explicit, it is still not enough. Recall that the goal is to
compare \grow\ to the action of $[F_{123},\eta]$ on the imaginary
part of scalar box functions. As discussed in section 2 and more
fully in appendix A, these imaginary parts are logarithms of
simple functions of the kinematical invariants. Once the collinear
operator is applied, it produces sums over different poles which
we loosely call ``simple fractions".

Therefore, we have to expand \grow\ in ``simple fractions". The
precise meaning of this will become clear as we do it.

Note that the overall coefficient of \grow\ is not relevant as it
is annihilated by $[F_{123},\eta]$, so we have to concentrate on
the terms inside the parenthesis. Moreover, $[F_{123},\eta]$ also
annihilates ${\langle 1 |P|4]}$ and ${\langle 3 |P|n]}$, so we
have to factor them out
\eqn\wrip{ [F_{123},\eta] C_{123} = {(t_1^{[3]})^3\over
\vev{1~2}\vev{2~3}[4~5][5~6]\ldots [n-1 ~ n]{\langle 1
|P|4]}{\langle 3 |P|n]}} \left( {[1~\eta]{\langle 3 |P|n]}\over
[n~1][1~2]} + {[3~\eta]{\langle 1 |P|4]}\over [2~3][3~4]} \right).
}

Recall that $P=p_1+p_2+p_3$. Using Schouten identity \schouten, we
find
\eqn\tefa{ \left( {[1~\eta]{\langle 3 |P|n]}\over [n~1][1~2]} +
{[3~\eta]{\langle 1 |P|4]}\over [2~3][3~4]}  \right) =
\vev{1~2}{{\langle 3|P|\eta ]} \over t_1^{[2]}} - \vev{2~3}{
{\langle 1|n|\eta ]}\over t_n^{[2]}} + \vev{2~3}{{\langle 1|P|\eta
]}\over t_2^{[2]}} - \vev{1~2}{{\langle 3|4|\eta ]}\over
t_3^{[2]}} }
Note that
\eqn\wiwus{\eqalign{ & \vev{1~2}{\langle 3|P|\eta ]}
=[F_{123},\eta]\left( t_1^{[2]}\right) , \quad \vev{2~3}{\langle
1|P|\eta ]} = [F_{123},\eta]\left( t_n^{[2]}\right), \cr &
\vev{2~3}{\langle 1|P|\eta ]} = [F_{123},\eta]\left(
t_2^{[2]}\right), \quad  \vev{1~2}{\langle 3|4|\eta ]} =
[F_{123},\eta]\left( t_3^{[2]}\right).}}

Finally, it is easy to identify each term in \tefa\ by using
\wiwus\ with the action of $[F_{123},\eta]$ on the logarithm of
$t_i^{[2]}$ for some $i$. Combining all logarithms into one we
find
\eqn\esim{ [F_{123},\eta] C_{123} = {(t_1^{[3]})^3\over
\vev{1~2}\vev{2~3}[4~5][5~6]\ldots [n-1 ~ n]{\langle 1
|P|4]}{\langle 3 |P|n]}}
  [F_{123},\eta] \log\left( {t_1^{[2]}t_2^{[2]}\over t_3^{[2]}t_n^{[2]}} \right). }

\subsec{Collinear Operator Acting On Box Functions}

In this section we compute the action of $ [F_{123},\eta]$ on
$C_{123}$ using the formula in terms of the imaginary part of box
functions \gola\
\eqn\gola{\eqalign{ C_{123} = {\rm Im}|_{t_1^{[3]}>0} & ~  \left(
b_4 F^{1m}_{n:4} + c_{2,2} F^{2m~e}_{n:2;2} + c_{3,1}
F^{2m~e}_{n:3;1} + c_{2,1} F^{2m~e}_{n:2;1} + d_{2,2}
 F^{2m~h}_{n:2;2} + d_{3,1} F^{2m~h}_{n:3;1} \right. \cr &  +
d_{n-5,6} F^{2m~h}_{n:n-5;6}+d_{n-4,5} F^{2m~h}_{n:n-4;5} +
\sum_{r'=2}^{n-5} g_{2,r',2} F^{3m}_{n:2:r';2} + \sum_{r'=2}^{n-6}
g_{3,r',1} F^{3m}_{n:3:r';1}\cr &  + \sum_{i=7}^{n-1}
g_{n-i-1,3,i} F^{3m}_{n:n-i-1:3;i} + \sum_{r=2}^{n-6}
g_{r,n-r-4,5} F^{3m}_{n:r:n-r-4;5} + \cr & \left. \sum_{r=2}^{n-4}
g_{r,n-r-3,4} F^{3m}_{n:r:n-r-3;4}  +
\sum_{r'=2}^{n-7}\sum_{r"=2}^{n-r'-5} f_{3,r',r'',1}
F^{4m}_{n:3:r':r'';1} \right).}}

Now we find the subtlety discussed at the end of section 2. It
turns out that $ [F_{123},\eta]$ annihilates each of the scalar
box functions in the following sum
\eqn\willy{\eqalign{ & c_{3,1} F^{2m~e}_{n:3;1} + d_{3,1}
F^{2m~h}_{n:3;1} + d_{n-5,6} F^{2m~h}_{n:n-5;6} +
\sum_{r'=2}^{n-6} g_{3,r',1} F^{3m}_{n:3:r';1}+ \sum_{i=7}^{n-1}
g_{n-i-1,3,i} F^{3m}_{n:n-i-1:3;i}\cr & + \sum_{r=2}^{n-6}
g_{r,n-r-4,5} F^{3m}_{n:r:n-r-4;5} +
\sum_{r'=2}^{n-7}\sum_{r"=2}^{n-r'-5} f_{3,r',r'',1}
F^{4m}_{n:3:r':r'';1}. }}
{}From their explicit form given in appendix B, it is easy to see
that each of them depends on gluons $1$, $2$, and $3$ only through
$p_1+p_2+p_3$. By the discussion that led to \schouten\ this is
condition is enough to ensure total annihilation by
$[F_{123},\eta]$.

Therefore, no information can be obtained about these coefficients
with this collinear operator. All this implies is that in order to
get the whole cut we have to consider the action of at least one
more collinear operator. We do this in section 3.3.

After removing the terms in \willy, we are left with
\eqn\letti{\eqalign{ [F_{123},\eta]C_{123} = &~[F_{123},\eta]\;
{\rm Im}|_{t_1^{[3]}>0}  \left( b_4 F^{1m}_{n:4} + c_{2,2}
F^{2m~e}_{n:2;2}  + c_{2,1} F^{2m~e}_{n:2;1} + d_{2,2}
 F^{2m~h}_{n:2;2}  \right. \cr & \left. + d_{n-4,5} F^{2m~h}_{n:n-4;5} +
\sum_{r'=2}^{n-5} g_{2,r',2} F^{3m}_{n:2:r';2} + \sum_{r=2}^{n-4}
g_{r,n-r-3,4} F^{3m}_{n:r:n-r-3;4} \right).}}

The imaginary parts of the box functions can be easily obtained as
explained in appendix A. Here we will concentrate first on those
scalar box function which produce a term that is not present in
\esim.

Consider for example,
\eqn\conan{  {\rm Im}|_{t_1^{[3]}>0}F^{2m~e}_{n:2;2} = \pi
\ln\left( 1- {t_2^{[2]}t_1^{[4]}\over t_1^{[3]}t_2^{[3]}} \right)
+ \ldots}
Upon acting with the collinear operator this produces a
contribution to \letti\ of the form
\eqn\ghelo{[F_{123},\eta]\ln ( t_2^{[2]}t_1^{[4]} -
t_1^{[3]}t_2^{[3]} )}
which is non zero and it is not present in \esim. Therefore we
conclude that $c_{2,2}=0$.

One might wonder whether there are other terms in
\letti\ that might cancel this contribution so that the equation
$c_{2,2}=0$ is be replaced by a relation between several
coefficients. As discussed in section 2, this is not the case, for
each box function has a ``unique signature".

Consider the limit when $t_1^{[3]}$ is large. Then $\ln
(t_2^{[2]}t_1^{[4]} - t_1^{[3]}t_2^{[3]})$ produces a term of the
form $\ln(-t_2^{[3]})$. Looking at \twoeasy\ we see that this is
the signature of $F^{2m~e}_{n:2;2}$ in the $t_1^{[3]}$ channel. In
other words, there is no other box function that could produce
such a term.

The same analysis can be repeated to find that $c_{2,1}
=g_{2,r',2} = g_{r,n-r-3,4} = 0$ for all $r$ and $r'$ in the sums
given in \letti.

Let us study in detail the imaginary parts of the three box
functions with nonzero coefficients remaining in \letti\
\eqn\imag{ \eqalign{{\rm Im}|_{t_1^{[3]}>0}& \; F^{1m}_{n:4} =
-\ln\left( 1- {t_1^{[3]}\over t_1^{[2]}} \right) -\ln\left( 1-
{t_1^{[3]}\over t_2^{[2]}} \right) ; \cr  {\rm Im}|_{t_1^{[3]}>0}&
\; F^{2m~h}_{n:2;2} = \ln\left( -{t_1^{[3]}\over t_n^{[2]}}
\right) + \ln\left( 1- {t_2^{[2]}\over t_1^{[3]}} \right) +
\ln\left( 1- {t_n^{[4]}\over t_1^{[3]}} \right); \cr  {\rm
Im}|_{t_1^{[3]}>0}& \; F^{2m~h}_{n:n-4;5} =  \ln\left(
-{t_1^{[3]}\over t_3^{[2]}} \right) + \ln\left( 1-
{t_5^{[n-4]}\over t_1^{[3]}} \right) + \ln\left( 1-
{t_1^{[2]}\over t_1^{[3]}} \right).}}
The problem at hand is to find $b_4, d_{2,2}$ and $d_{n-4,5}$ such
that \letti\ equals \esim. More explicitly, we need
\eqn\gop{\eqalign{ & [F_{123},\eta]\; {\rm Im}|_{t_1^{[3]}>0}
\left( b_4 F^{1m}_{n:4}  + d_{2,2}
 F^{2m~h}_{n:2;2} + d_{n-4,5} F^{2m~h}_{n:n-4;5}\right) =\cr & {(t_1^{[3]})^3\over
\vev{1~2}\vev{2~3}[4~5][5~6]\ldots [n-1 ~ n]{\langle 1
|P|4]}{\langle 3 |P|n]}}
  [F_{123},\eta] \ln\left( {t_1^{[2]}t_2^{[2]}\over t_3^{[2]}t_n^{[2]}}
  \right).}}

First note that the imaginary part of $F^{1m}_{n:4}$ is the only
one that contains the term $\ln (t_1^{[2]}t_2^{[2]})$, again by
looking at \oneM\ we find that this is the signature of
$F^{1m}_{n:4}$ in this channel. On the other hand,
$F^{2m~h}_{n:2;2}$ is the only function that produces $\ln (-
t_n^{[2]})$ and finally, $\ln(-t_3^{[2]})$ is unique to the
imaginary part of $F^{2m~h}_{n:n-4;5}$. This implies that
\eqn\juki{ b_4 = d_{2,2} = d_{n-4,5} = {(t_1^{[3]})^3\over
\vev{1~2}\vev{2~3}[4~5][5~6]\ldots [n-1 ~ n]{\langle 1
|P|4]}{\langle 3 |P|n]}}. }
Indeed, setting all coefficients equal and realizing that
\eqn\wuwo{[F_{123},\eta]\;\ln\left( 1- {t_n^{[4]}\over t_1^{[3]}}
\right) = 0, \qquad  [F_{123},\eta]\;\ln\left( 1- {t_5^{[2]}\over
t_1^{[3]}} \right) = 0, }
one can easily check that
\eqn\qoqi{ [F_{123},\eta]\; {\rm Im}|_{t_1^{[3]}>0} \left(
F^{1m}_{n:4}  +
 F^{2m~h}_{n:2;2} +  F^{2m~h}_{n:n-4;5}\right) = [F_{123},\eta] \ln\left( {t_1^{[2]}t_2^{[2]}\over t_3^{[2]}t_n^{[2]}}
  \right).}

\subsec{Computing The Remaining Coefficients}

As mentioned above, the coefficients in \willy\ remain unknown and
have to be determined by the action of a different collinear
operator. From looking at \hiho\ we see that no other set of
gluons is manifestly localized on a line. To see this note that
the second tree-level amplitude  in \hiho\ is a mostly minus MHV
amplitude and therefore it has gluons localized on a degree $n-4$
curve.

The solution to this problem is clear, consider $C_{123}^\dagger$
instead of $C_{123}$. Then the mostly minus tree-level amplitude
becomes a mostly plus MHV amplitude which is localized on a line.
Now we can consider the action of $[F_{4in},\eta]$, with $i$
taking any value from $5$ to $n-1$, on $C_{123}^\dagger$.

It turns out that $[F_{4in},\eta]$ only annihilates
$F^{1m}_{n:4}$. Therefore, this analysis does not provide
information about $b_4^\dagger$. However, $b_4$ is already known
\juki.

Let us start again by writing $C_{123}^{\dagger}$ explicitly
\eqn\usso{\eqalign{ C_{123}^\dagger & = \int d\mu A^{\rm
tree}((-\ell_1)^+,1^-,2^-,3^-,(-\ell_2)^+)A^{\rm
tree}(\ell_2^-,4^+,5^+,\ldots ,n^+,\ell_1^-) \cr & = -{1\over
[1~2][2~3]\vev{4~5}\vev{5~6}\ldots \vev{n-1 ~ n}}\int d\mu
{[\ell_1~\ell_2]^3\vev{\ell_1~\ell_2 }^3\over [\ell_1~
1][3~\ell_2]\vev{\ell_2~4}\vev{n~\ell_1} } \cr & =
-{(t_1^{[3]})^3\over [1~2][2~3]\vev{4~5}\vev{5~6}\ldots \vev{n-1 ~
n}} \int d\mu {1\over [\ell_1~
1][3~\ell_2]\vev{\ell_2~4}\vev{n~\ell_1}} }.}

Following exactly the same steps as for $[F_{123},\eta]C_{123}$,
we find
\eqn\reman{\eqalign{& [F_{4in},\eta]C_{123}^\dagger =\cr &
{(t_1^{[3]})^3\over [1~2][2~3]\vev{4~5}\vev{5~6}\ldots
\vev{n-1~n}{\langle 4|P|1]}{\langle n|P|3]}}\left(
\vev{4~i}{[n~\eta]{\langle 4|P|1]}\over [n~1]{\langle 4|P|n]}} +
\vev{i~n}{[4~\eta]{\langle n|P|3]}\over [4~3]{\langle
n|P|4]}}\right).}}

Using that $P=p_1+p_2+p_3$ and several Schouten's identity we can
expand the term in parenthesis in ``simple fractions"
\eqn\fracy{ \vev{4~i}{{\langle 4|P|\eta ]}\over {\langle 4|P|n]}}
-\vev{4~i}{[\eta~1]\over [n~1]} + \vev{i~n}{{\langle n|P|\eta
]}\over {\langle n|P|4]}} -\vev{i~n}{[\eta~3]\over [4~3]}.}

In this form, it is easy to identify \fracy\ with
\eqn\joker{ [F_{4in},\eta]\ln\left( {{\langle 4|P|n]}{\langle
n|P|4]} \over [n~1][4~3]} \right) = -  [F_{4in},\eta]\ln\left(
{t_n^{[2]}t_3^{[2]} \over t_n^{[4]}t_1^{[4]} -t_1^{[3]}t_n^{[5]}}
\right).}

Combining \joker\ with \reman\ we finally find
\eqn\yeye{ [F_{4in},\eta]C_{123}^\dagger  = {(t_1^{[3]})^3\over
[1~2][2~3]\vev{4~5}\vev{5~6}\ldots \vev{n-1~n}{\langle
4|P|1]}{\langle n|P|3]}}  [F_{4in},\eta]\ln\left(
{t_n^{[2]}t_3^{[2]} \over t_n^{[4]}t_1^{[4]} -t_1^{[3]}t_n^{[5]}}
\right). }

As in the analysis of section 3.2, we have to compare \yeye\ to
the action of $[F_{4in},\eta]$ on the imaginary part of the box
functions in \gola. Recall that \yeye\ is supposed to provide
information about the coefficients that remained unknown after the
study of $[F_{123},\eta]C_{123}$, i.e., the coefficients in
\willy.

By computing the imaginary part of the box functions we see that
in order to reproduce \yeye\ we have to impose that $d_{3,1}$,
$d_{n-5,6}$, $g_{3,r',1}$, $g_{n-i-1,3,i}$, $g_{r,n-r-4,5}$ and
$f_{3,r',r'',1}$ all vanish. Thus, the only coefficient left in
\willy\ is $c_{3,1}$.

Let us write down the imaginary part of $F^{2m~e}_{n:3;1}$
explicitly,
\eqn\iman{  {\rm Im}|_{t_1^{[3]}>0}~ F^{2m~e}_{n:3;1} = -\ln
\left( 1- {t_1^{[3]}\over t_n^{[4]}} \right) - \ln \left( 1-
{t_1^{[3]}\over t_1^{[4]}} \right) + \ln \left( 1- {t_1^{[3]}
t^{[5]}_n\over t_n^{[4]}t_1^{[4]}} \right). }
{}From this we can see that this contributes a factor of $\ln (
t_n^{[4]}t_1^{[4]} -t_1^{[3]}t_n^{[5]})$ and therefore
$c_{3,1}^\dagger$ must equal the overall factor in \yeye.

By comparing the overall factor of \gop\ with that of \yeye\ we
find that one is the complex conjugate of the other. Therefore,
using \juki\ we find
\eqn\fuu{\eqalign{ [F_{4in},\eta]C_{123}^\dagger =&~
{(t_1^{[3]})^3\over [1~2][2~3]\vev{4~5}\vev{5~6}\ldots
\vev{n-1~n}{\langle 4|P|1]}{\langle n|P|3]}}\times \cr &
[F_{4in},\eta] {\rm Im}|_{t_1^{[3]}>0} \left(F_{n:2;2}^{2m~h} +
F^{2m~h}_{n:n-4;5} + F^{2m~e}_{m:3;1} \right).}}

We now explicitly check that \fuu\ equals \yeye. Even though this
is guaranteed to work, it is still interesting to see the
interplay between the different imaginary parts.

Using the explicit formulas for the imaginary parts of the box
functions involved in \fuu\ given in \imag\ and \iman\ we find
that \fuu\ equals
\eqn\goal{ \eqalign{ [F_{4in},\eta]C_{123}^\dagger &~ =
{(t_1^{[3]})^3\over [1~2][2~3]\vev{4~5}\vev{5~6}\ldots
\vev{n-1~n}{\langle 4|P|1]}{\langle n|P|3]}}\times \cr &
[F_{4in},\eta] {\rm Im}|_{t_1^{[3]}>0} \left( \ln\left(
{t_n^{[2]}t_3^{[2]} \over t_n^{[4]}t_1^{[4]} -t_1^{[3]}t_n^{[5]}}
\right) + \ln \left( 1-{t_2^{[2]}\over t_1^{[3]}} \right) + \ln
\left( 1-{t_1^{[2]}\over t_1^{[3]}} \right) \right).}}
Note that the last two terms in \goal\ are trivially annihilated
by $[F_{4in},\eta]$ providing the desired result.

\subsec{Summary Of Results}

Let us collect all the results we have obtained in order to write
down the final formula for the cut $C_{123}$.

Recall that the cut was given by \gola\
\eqn\wimp{\eqalign{ C_{123} = {\rm Im}|_{t_1^{[3]}>0} & ~  \left(
b_4 F^{1m}_{n:4} + c_{2,2} F^{2m~e}_{n:2;2} + c_{3,1}
F^{2m~e}_{n:3;1} + c_{2,1} F^{2m~e}_{n:2;1} + d_{2,2}
 F^{2m~h}_{n:2;2} + d_{3,1} F^{2m~h}_{n:3;1} \right. \cr &  +
d_{n-5,6} F^{2m~h}_{n:n-5;6}+d_{n-4,5} F^{2m~h}_{n:n-4;5} +
\sum_{r'=2}^{n-5} g_{2,r',2} F^{3m}_{n:2:r';2} + \sum_{r'=2}^{n-6}
g_{3,r',1} F^{3m}_{n:3:r';1}\cr &  + \sum_{i=7}^{n-1}
g_{n-i-1,3,i} F^{3m}_{n:n-i-1:3;i} + \sum_{r=2}^{n-6}
g_{r,n-r-4,5} F^{3m}_{n:r:n-r-4;5} + \cr & \left. \sum_{r=2}^{n-4}
g_{r,n-r-3,4} F^{3m}_{n:r:n-r-3;4}  +
\sum_{r'=2}^{n-7}\sum_{r"=2}^{n-r'-5} f_{3,r',r'',1}
F^{4m}_{n:3:r':r'';1} \right).}}

It is convenient to define
\eqn\defB{ {\cal B}_n = {(t_1^{[3]})^3\over
\vev{1~2}\vev{2~3}[4~5][5~6]\ldots [n-1 ~ n]{\langle 1
|P|4]}{\langle 3 |P|n]}}. }
{}From the analysis of $[F_{123},\eta]C_{123}$ we found
\eqn\suzi{\eqalign{ & b_4 = d_{2,2} = d_{n-4,5} = {\cal B}_n, \cr
& c_{2,2} = c_{2,1} =g_{2,r',2} =g_{r,n-r-3,4} = 0 .}}
And from the analysis of $[F_{4in},\eta]C_{123}^\dagger$ we found
\eqn\yizi{ \eqalign{ & c_{3,1} = {\cal B}_n, \cr &
d_{3,1}=d_{n-5,6}=g_{3,r',1}=g_{n-i-1,3,i}=g_{r,n-r-4,5}
=f_{3,r',r'',1}= 0. } }
Putting together \suzi\ and \yizi\ we have all coefficients
appearing in \wimp. The final result for the cut is then
\eqn\gromp{C_{123} = {\cal B}_n ~ {\rm Im}|_{t_1^{[3]}>0}\left(
F^{1m}_{n:4} + F^{2m~e}_{n:3;1} + F^{2m~h}_{n:2;2} +
F^{2m~h}_{n:n-4;5} \right).}

\newsec{Consistency Checks}

In this section, we study several consistency checks that \gromp\
must satisfy. The first is the requirement that the cut must be
free of the IR divergencies that show up in the scalar box
functions. The others are just the comparison of \gromp\ with the
cuts of known amplitudes, i.e., $n=4,5$ and $6$.

\subsec{Finiteness Of The Cut}

All box functions, except for the four-mass, are infrared
divergent. This means that they have to be evaluated using some
regularization procedure. The explicit formulas in appendix A are
in the dimensional regularization scheme. On the other hand, the
$(1,2,3)$-cut we have computed has to be finite. The reason is
that, as mentioned in section 2, the integration region is
compact; it is a two-sphere. Therefore, the divergencies of the
imaginary part of each of the box functions appearing in \gromp\
must cancel out.

The infrared divergent structure of each of these function is
given by
\eqn\hios{\eqalign{  F^{1m}_{n:4}|_{\rm IR} = &~ -{1\over
\epsilon^2}\left[ (-t_{1}^{[2]})^{-\epsilon}
+(-t_{2}^{[2]})^{-\epsilon} -(-t_{1}^{[3]})^{-\epsilon} \right];
\cr  F^{2m~e}_{n:3;1}|_{\rm IR} = &~ -{1\over \epsilon^2}\left[
(-t_{n}^{[4]})^{-\epsilon} +(-t_{1}^{[4]})^{-\epsilon}
-(-t_{1}^{[3]})^{-\epsilon} - (-t_{n}^{[5]})^{-\epsilon} \right];
\cr   F^{2m~h}_{n:2;2}|_{\rm IR} = &~  -{1\over \epsilon^2}\left[
\half (-t_{n}^{[2]})^{-\epsilon} +(-t_{1}^{[3]})^{-\epsilon}
-\half (-t_{1}^{[2]})^{-\epsilon} - \half
(-t_{n}^{[4]})^{-\epsilon} \right]; \cr F^{2m~h}_{n:n-4;5}|_{\rm
IR} = &~
 -{1\over \epsilon^2}\left[
\half (-t_{3}^{[2]})^{-\epsilon} +(-t_{4}^{[n-3]})^{-\epsilon}
-\half (-t_{5}^{[n-4]})^{-\epsilon} - \half
(-t_{1}^{[2]})^{-\epsilon} \right] .} }

It is clear that the only term in each of these functions that
develops an imaginary part is $(-t_{1}^{[3]})^{-\epsilon}$. Note
that in the sum they cancel out. To see this more clearly, recall
that by momentum conservation $t_{4}^{[n-3]}=t_1^{[3]}$.

\subsec{Four- and Five-Gluon Amplitudes}

Taking $n=4$ we find that ${\cal B}_n$ is trivially zero. To see
this note that $t_1^{[3]} = p_4^2 = 0$. The zeroes in the
denominator are cancelled by some of the three powers of
$t_1^{[3]}$ in the numerator. This is consistent with the fact
that the amplitude $A_{4:1}(1^+,2^+,3^+,4^-)$ is exactly zero.

More interesting is the case when $n=5$. This amplitude was first
computed in \BernMQ. Setting $n=5$ in the definition of ${\cal
B}_n$ we find that
\eqn\nfive{ {\cal B}_5 = {\vev{4~5}^3\over
\vev{5~1}\vev{1~2}\vev{2~3}\vev{3~4}}.}
This is the-tree level MHV amplitude $A^{\rm
tree}(1^+,2^+,3^+,4^-,5^-)$. Also note that for $n=5$ the last
three box functions in \gromp\ naturally descend to one-mass
scalar box functions as follows: $F^{2m~e}_{5:3;1}=0$,
$F^{2m~h}_{5:2;2} = F^{1m}_{5:2}$ and $F^{2m~h}_{5:1;5} =
F^{1m}_{5;1}$. This gives
\eqn\ampF{ C_{123} = A^{\rm tree}(1^+,2^+,3^+,4^-,5^-)~{\rm
Im}|_{t_1^{[3]}>0}\left( F^{1m}_{5:4} +
F^{1m}_{5:2}+F^{1m}_{5;1}\right)}
which is the right answer for the cut.

\subsec{Six-Gluon Amplitude}

The one-loop non-MHV six gluon partial amplitude
$A_{6;1}^{\N=4}(1^+,2^+,3^+,4^-,5^-,6^-)$ was first computed by
Bern et.al. in \BernCG. They computed $C_{123}$ by different
methods to the ones presented here and obtained
\eqn\quill{ C_{123} = B_0 ~{\rm Im}|_{t_1^{[3]}>0}\left(
F^{1m}_{6:1} + F^{1m}_{6:4}+F^{2m~h}_{6:2;2}+F^{2m~h}_{6:2;5}
\right).}
where
\eqn\bsix{ B_0 =
{([1~2]\vev{2~4}+[1~3]\vev{3~4})([3~1]\vev{1~6}+[3~2]\vev{2~6})(t_1^{[3]})^3\over
\vev{1~2}\vev{2~3}[4~5][5~6](t_1^{[3]}t_{3}^{[3]}-t_{1}^{[2]}t_{4}^{[2]})(t_1^{[3]}t_{2}^{[3]}-t_{2}^{[2]}t_{5}^{[2]})}.
}

In order to show that \quill\ is equal to \gromp\ with $n=6$ all
we have to do is to realize that
\eqn\nulli{ \eqalign{ F^{2m~e}_{6:3;1} = &~ F^{1m}_{6:1} \cr
t_1^{[3]}t_{3}^{[3]}-t_{1}^{[2]}t_{4}^{[5]} =&~ {\langle
6|P|3]}{\langle 3|P|6]} =
 ([3~1]\vev{1~6}+[3~2]\vev{2~6}){\langle 3|P|6]} \cr
t_1^{[3]}t_{2}^{[3]}-t_{2}^{[2]}t_{5}^{[2]} =&~ {\langle
4|P|1]}{\langle 1|P|4]} =([1~2]\vev{2~4}+[1~3]\vev{3~4}){\langle
1|P|4]}. } }

This makes it manifest that $B_0$ in \bsix\ is equal to our ${\cal
B}_6$ in \defB.

Something worth mentioning is that all other independent cuts are
given in terms of $C_{123}$ \BernCG. In our notation, $C_{234}$ is
given as follows,
\eqn\remy{ C_{234} = \left( {[2~3]\vev{5~6}\over
t_{234}}\right)^4\times \left.\left[ C^{\dagger}_{123}
\right]\right|_{j\rightarrow j+1}  + \left( {{\langle 1|P|4]}
\over t_{234}}\right)^4\times \left.\left[ C_{123}
\right]\right|_{j\rightarrow j+1} }
The remaining cut, i.e., $C_{345}$ can be obtained from \remy\ by
shifting the labels and conjugation.

Knowing all cuts implies that we know all coefficients in the
amplitude and thus the amplitude itself. For the explicit form of
the amplitude see \BernCG.

\newsec{Discussion}

We have shown that certain unitarity cuts can be determined in
terms of scalar box functions in a simple way. The class of cuts
for which the method proposed here works involves all the cuts of
two interesting series of amplitudes. Namely, the n-gluon one-loop
MHV amplitudes and the n-gluon one-loop next-to-MHV amplitudes
with three {\it consecutive} plus helicity gluons and the rest
negative. We have checked that our method reproduces correctly all
MHV amplitudes. We have not included the computation here because
it is very similar to that of our main example.

We have explicitly computed one of the cuts in the next-to-MHV
series to show how efficient this method can be. It is important
to remark that for $n=6$ the $(1,2,3)$-cut was computed about ten
years ago in a pioneering work by Bern, et.al. \BernCG. That
computation uses very powerful reduction techniques \BernKR.
However, it leads to quite complicated formulas before it can be
put in the final simple form. Notably, the six-gluon amplitude
computed in \BernCG\ is actually the only one-loop non-MHV
amplitude known to date.

Although we have explicitly computed only one of the cuts of the
next-to-MHV series, the computation of the remaining cuts should
be within reasonable reach. This implies that the computation of
the corresponding series of amplitudes is also within reach.

Even more interesting is the possibility of extending this method
to all possible cuts. Clearly, this would have to involve more
than just collinear operators since, in general, both tree-level
amplitudes appearing in the cut will be non-MHV. However, in
\CachazoKJ\ a prescription was given to compute all tree-level
amplitudes in terms of MHV diagrams. These diagrams are made out
of MHV amplitudes continued off-shell and connected by Feynman
propagators. In twistor space this corresponds to configurations
where gluons are localized on unions of lines. Moreover, from the
viewpoint of the differential operators (see section 2 of
\CachazoZB\ for more details), these lines intersect to form
quivers or trees. So, it is conceivable that by combining
collinear operators with coplanar operators (that tests whether
four gluons are contained in a common plane in twistor space), one
could compute all cuts. Although coplanar operators have not been
discussed in the context of the holomorphic anomaly, they are
indeed affected by it.

\bigskip
\bigskip
\centerline{\bf Acknowledgements}

It is a pleasure to thanks R. Britto, A. De Freitas, P. Svr\v cek
and E. Witten for helpful discussions. This work was supported in
part by the Martin A. and Helen Chooljian Membership at the
Institute for Advanced Study and by DOE grant DE-FG02-90ER40542.

\appendix{A}{Scalar Box Functions And Monodromies}

Scalar Box functions are a set of functions constructed from the
scalar box integrals. The latter form a complete set of the
possible integrals that can appear in a Feynman diagrammatic
computation of one-loop amplitudes in ${\cal N}=4$ gauge
theory.\foot{After Passarino-Veltman reduction formulas are
applied.}

These integrals are known as the scalar box integrals because they
would arise in a one-loop computation of a scalar field theory
with four internal propagators.

\ifig\pipo{Scalar box integrals used in the definition of scalar
box functions: (a) One-mass $F^{1m}_{n:i}$. (b) Two-mass ``easy"
$F^{2m~e}_{n:r;i}$. (c) Two-mass ``hard" $F^{2m~h}_{n:r;i}$. (d)
Three-mass $F^{3m}_{n:r:r';i}$. (e) Four-mass
$F^{4m}_{n:r:r':r";i}$.}
{\epsfxsize=0.85\hsize\epsfbox{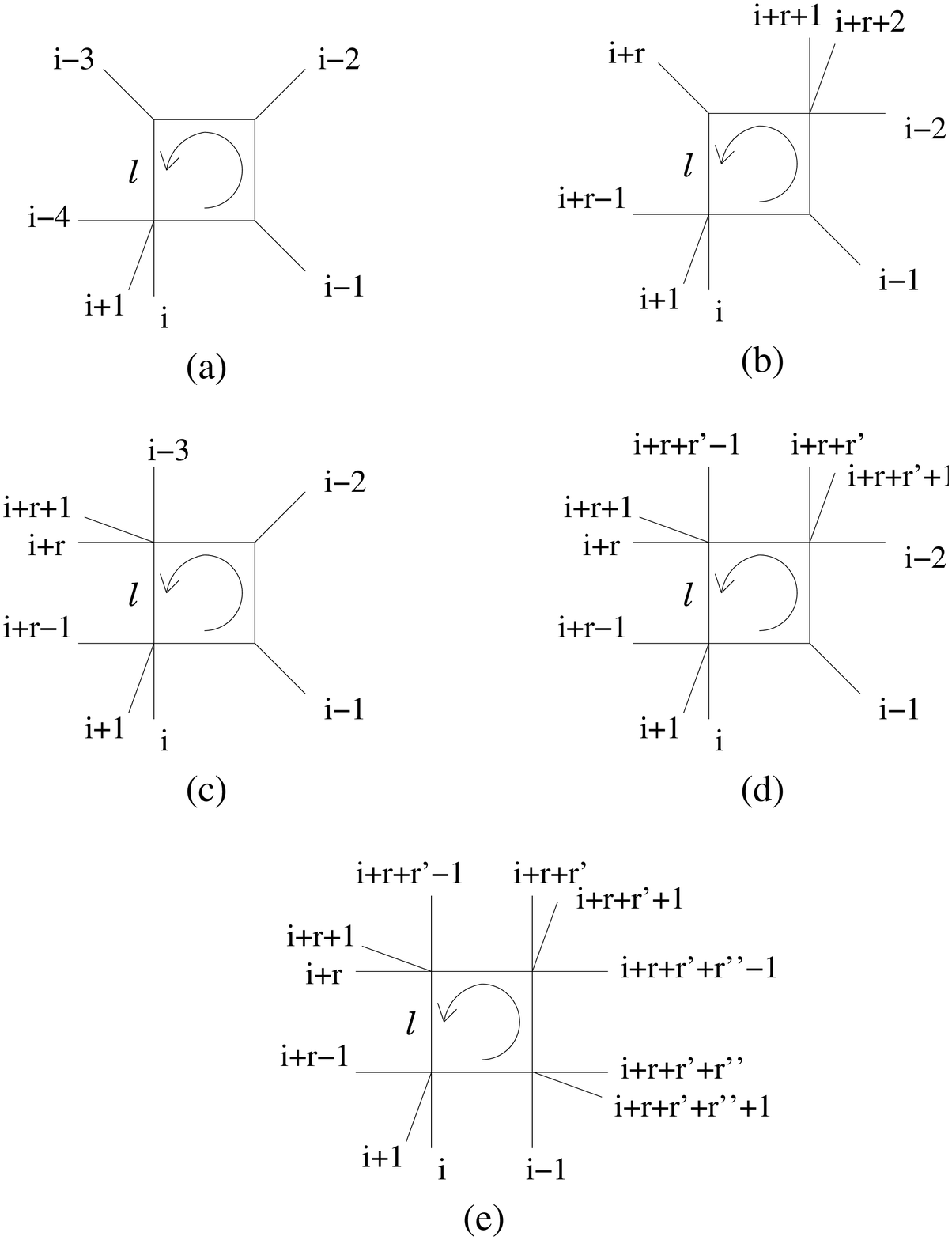}}

The scalar box integral is defined as follows:
\eqn\bbo{ I_4 = -i(4\pi)^{2-\epsilon} \int
{d^{4-2\epsilon}\ell\over (2\pi)^{4-2\epsilon}} {1\over \ell^2
(\ell-K_1)^2(\ell-K_1-K_2)^2(\ell+K_4)^2}. }
The incoming external momenta at each of the vertices are
$K_1,K_2,K_3,K_4$. The labels are given in consecutive order
following the loop. Momentum conservation implies that
$K_1+K_2+K_3+K_4 = 0$ and this is why \bbo\ only depends on three
momenta.

In the computation of one-loop $\N=4$ amplitudes, each of the
external momenta becomes a sum of the momenta of external gluons.
The evaluation of the integral \bbo\ varies in complexity
depending upon the number of legs that have $K_i^2=0$, i.e., $K_i$
is just a single gluon momentum. The convention is to label the
integrals according to the number of $K_i^2\neq 0$, the so-called
``massive legs". In general, the number of massive legs goes from
1 to 4. For two masses there are two inequivalent choices of the
massive legs and are called ``easy" and ``hard". The names refer
to how difficult the evaluation of \bbo\ is compared to the other.
The different possibilities, as shown in \pipo, are:
\eqn\poqo{ I^{1m}_{4:i},~ I^{2m~e}_{n:r;i},~ I^{2m~h}_{4:r;i},~
I^{3m}_{4:r:r';i},~ I^{4m}_{4:r:r':r'';i}.}

It turns out to be convenient to introduce the scalar box
functions. These are defined as follows
\eqn\deki{ \eqalign{ & I^{1m}_{4:i} = -2{F^{1m}_{n:i}\over
t_{i-3}^{[2]}t_{i-2}^{[2]}}, \quad I^{2m~e}_{4:r;i} =
-2{F^{2m~e}_{n:r;i}\over
t_{i-1}^{[r+1]}t_i^{[r+1]}-t_i^{[r]}t_{i+r+1}^{[n-r-2]}}, \quad
I^{2m~h}_{4:r;i} = -2{F^{2m~h}_{n:r;i}\over
t_{i-1}^{[2]}t_{i-1}^{[r+1]} }, \cr & I^{3m}_{4:r:r';i} =
-2{F^{3m}_{n:r:r';i}\over
t_{i-1}^{[r+1]}t_i^{[r+r']}-t_i^{[r]}t_{i+r+r'}^{[n-r-r'-1]} },
\quad I^{4m}_{4:r:r':r";i} = -2{F^{4m}_{n:r:r':r";i}\over
t^{[r+r']}_{i}t^{[r'+r"]}_{i+r}\rho}, } }
where
\eqn\spell{ \rho = \sqrt{1-2\lambda_1-2\lambda_2+\lambda_1^2 -
2\lambda_1\lambda_2+\lambda_2^2},}
and
\eqn\andi{ \lambda_1={t_i^{[r]}t_{i+r+r'}^{[r"]}\over
t_i^{[r+r']}t_{i+r}^{r'+r"}}, \quad
\lambda_2={t_{i+r}^{[r']}t_{i+r+r'+r"}^{[n-r-r'-r"]}\over
t_i^{[r+r']}t_{i+r}^{r'+r"}}.}
We have set to one a factor usually denote by $r_{\Gamma}$. The
reason being that we are only dealing with cuts that are finite
and $r_{\Gamma}$ goes to one as $\epsilon$, the dimensional
regularization parameter, goes to 0.

\subsec{Monodromy Analysis}

The box functions have branch cuts and monodromies. These
monodromies can be computed as the imaginary part of the box
function in an appropriate kinematical region. Here we provide a
general way of computing the monodromies in all channels of
interest and some explicit examples.

Let us start by writing down the explicit form of the box
functions,
\eqn\expi{\eqalign{ F^{1m}_{n:i} = & ~ -{1\over \epsilon^2}\left[
(-t_{i-3}^{[2]})^{-\epsilon} +(-t_{i-2}^{[2]})^{-\epsilon}
-(-t_{i-3}^{[3]})^{-\epsilon} \right] \cr & + {\rm Li}_2 \left( 1-
{t_{i-3}^{[3]}\over t_{i-3}^{[2]}} \right) + {\rm Li}_2 \left( 1-
{t_{i-3}^{[3]}\over t_{i-2}^{[2]}} \right) + {1\over 2}\ln^2\left(
{t_{i-3}^{[2]}\over t_{i-2}^{[2]}} \right) + {\pi^2\over 6},}}
\eqn\exop{\eqalign{ F^{2m~e}_{n:r;i} = & ~ -{1\over
\epsilon^2}\left[ (-t_{i-1}^{[r+1]})^{-\epsilon}
+(-t_{i}^{[r+1]})^{-\epsilon} -(-t_{i}^{[r]})^{-\epsilon} -
(-t_{i-1}^{[r+2]})^{-\epsilon} \right] \cr & + {\rm Li}_2 \left(
1- {t_{i}^{[r]}\over t_{i-1}^{[r+1]}} \right) + {\rm Li}_2 \left(
1- {t_{i}^{[r]}\over t_{i}^{[r+1]}} \right) +{\rm Li}_2 \left( 1-
{t_{i-1}^{[r+2]}\over t_{i-1}^{[r+1]}} \right) \cr & + {\rm Li}_2
\left( 1- {t_{i-1}^{[r+2]}\over t_{i}^{[r+1]}} \right) - {\rm
Li}_2 \left( 1- {t_{i}^{[r]}t_{i-1}^{[r+2]}\over
t_{i-1}^{[r+1]}t_{i}^{[r+1]}} \right) + {1\over 2}\ln^2\left(
{t_{i-1}^{[r+1]}\over t_{i}^{[r+1]}} \right),}}
\eqn\uxop{\eqalign{ F^{2m~h}_{n:r;i} = & ~ -{1\over
\epsilon^2}\left[ (-t_{i-2}^{[2]})^{-\epsilon}
+(-t_{i-1}^{[r+1]})^{-\epsilon} -(-t_{i}^{[r]})^{-\epsilon} -
(-t_{i-2}^{[r+2]})^{-\epsilon} \right] \cr & -{1\over
2\epsilon^2}{(-t_{i}^{[r]})^{-\epsilon}(-t_{i-2}^{[r+2]})^{-\epsilon}
\over (-t_{i-2}^{[2]})^{-\epsilon}}  + {1\over 2}\ln^2 \left(
{t_{i-2}^{[2]}\over t_{i-1}^{[r+1]}} \right) \cr & +  {\rm Li}_2
\left( 1- {t_{i}^{[r]}\over t_{i-1}^{[r+1]}} \right) + {\rm Li}_2
\left( 1- {t_{i-2}^{[r+2]}\over t_{i-1}^{[r+1]}} \right), }}
\eqn\uliz{\eqalign{ F^{3m}_{n:r:r';i} = & ~ -{1\over
\epsilon^2}\left[ (-t_{i-1}^{[r+1]})^{-\epsilon}
+(-t_{i}^{[r+r']})^{-\epsilon} -(-t_{i}^{[r]})^{-\epsilon} -
(-t_{i+r}^{[r']})^{-\epsilon} - (-t_{i-1}^{[r+r'+1]})^{-\epsilon}
\right] \cr & -{1\over
2\epsilon^2}{(-t_{i}^{[r]})^{-\epsilon}(-t_{i+r}^{[r']})^{-\epsilon}
\over (-t_{i}^{[r+r']})^{-\epsilon}} -{1\over
2\epsilon^2}{(-t_{i+r}^{[r']})^{-\epsilon}(-t_{i-1}^{[r+r'+1]})^{-\epsilon}
\over (-t_{i-1}^{[r+1]})^{-\epsilon}} + {1\over 2}\ln^2 \left(
{t_{i-1}^{[r+1]}\over t_{i}^{[r+r']}} \right) \cr & +  {\rm Li}_2
\left( 1- {t_{i}^{[r]}\over t_{i-1}^{[r+1]}} \right) + {\rm Li}_2
\left( 1- {t_{i-1}^{[r+r'+1]}\over t_{i}^{[r+r']}} \right) - {\rm
Li}_2 \left( 1- {t_i^{[r]}t_{i-1}^{[r+r'+1]}\over
t_{i-1}^{[r+1]}t_{i}^{[r+r']}} \right), }}
\eqn\woop{\eqalign{ F^{4m}_{n:r:r':r";i} = & ~ {1\over 2}\left(
-{\rm Li}_2 ( K_{-++} ) + {\rm Li}_2 ( K_{-+-}) - {\rm Li}_2
\left( -{1\over \lambda_1}K_{---} \right)+ {\rm Li}_2 \left(
-{1\over \lambda_1}K_{--+} \right) \right. \cr & ~ \left. -
{1\over 2}\ln \left( {\lambda_1\over \lambda_2^2} \right)
\ln\left({K_{+-+}\over K_{+--}} \right)\right)}}
where
\eqn\weki{ K_{\epsilon_1, \epsilon_2, \epsilon_3} = {1\over
2}\left( 1+ \epsilon_1 \lambda_1 + \epsilon_2 \lambda_2 +
\epsilon_3 \rho \right). }

In these formulas, ${\rm Li}_2(x)$ denotes the dilogarithm
function as defined by Euler, i.e., ${\rm Li}_2 = -\int_0^x \ln
(1-z)dz/z$.

We now turn to the study of the monodromies. The discussion that
follows is applicable for the first three sets of scalars box
functions. The four-mass scalar box function is more complicated
and the analysis of its analytic structure is done at the end of
this appendix.

Suppose that we want to compute the monodromy around the branch
cut in the $t_i^{[r]}$ channel. The appropriate kinematical regime
is that in which $t_i^{[r]}$ is positive and all other invariants
are negative. The monodromy is related to the imaginary part of
the function in this special regime\foot{The sign depends on the
direction the cut is crossed.}. We can compute the imaginary part
of the dilogarithms by using Euler's identity,
\eqn\euler{ {\rm Li}_2 (1-z) = {\rm Li}_2(z) + {\pi^2\over 6} -
\ln z \ln (1-z).}
The dilogarithm ${\rm Li}_2(z)$ is real for $z<1$, therefore, for
any function $g$ which is negative in the kinematical regime of
interest we have
\eqn\monodr{\eqalign{  \Delta {\rm Li}_2\left( 1 - g t_i^{[r]}
\right) = & ~ {\Im}|_{t_i^{[r]}>0} {\rm Li}_2\left( 1 - g
t_i^{[r]} \right) = - \pi \ln \left( 1- g t_i^{[r]} \right), \cr
\Delta {\rm Li}_2\left( 1 - {g\over t_i^{[r]}} \right) = & ~ -
{\Im}|_{t_i^{[r]}>0} {\rm Li}_2\left( 1 - {g \over t_i^{[r]}}
\right) = \pi \ln \left( 1- {g\over t_i^{[r]}} \right), \cr \Delta
\ln^2\left( {g\over t_i^{[r]}} \right) =  & ~
{\Im}|_{t_i^{[r]}>0}\ln^2\left( {g\over t_i^{[r]}} \right) = 2 \pi
\ln \left( {g\over t_i^{[r]}} \right) .}}
These formulas cover all the monodromies of the finite part of the
box functions.

The infrared divergent terms of the form
\eqn\irdi{ {1\over \epsilon^2}(-t_{i}^{[r]})^{-\epsilon} }
also develop an imaginary part
\eqn\imdiv{ {\Im}|_{t_i^{[r]}>0}\left( {1\over
\epsilon^2}(-t_{i}^{[r]})^{-\epsilon} \right) = {2\pi \over
\epsilon}\ln t_i^{[r]} + {\cal O}(\epsilon^0).}
In most of the paper we ignore this contribution but in section
3.1 it is explicitly shown to cancel out.

Finally, let us consider some useful examples of the monodromies
around the $t_1^{[3]}$ cut:

\item{1.} Consider the ``one-mass" box function \expi. The only
such function with a cut in this channel\foot{For $n=6$ there are
two possibilities: $F^{1m}_{6:1}$ and $F^{1m}_{6:4}$. For $n=5$
there are three: $F^{1m}_{5:1}$, $F^{1m}_{5:4}$, and
$F^{1m}_{5:2}$.} is $F^{1m}_{n:4}$,
\eqn\cutonemass{ \Delta F^{1m}_{n:4} = - {\rm ln} \left( 1-
{t_{1}^{[3]}\over t_{1}^{[2]}} \right) - {\rm ln} \left( 1-
{t_{1}^{[3]}\over t_{2}^{[2]}} \right).}

\item{2.} Consider the ``two-mass-easy" box function \exop. There
are three such functions with a cut in this channel:
$F^{2m~e}_{n:2;2}$,$F^{2m~e}_{n:3;1}$, and $F^{2m~e}_{n:2;1}$. let
us study the first,
\eqn\cuttwomass{ \Delta F^{2m~e}_{n:2;2} = {\rm ln} \left( 1-
{t_{2}^{[2]}\over t_{1}^{[3]}} \right) +{\rm ln} \left( 1-
{t_{1}^{[4]}\over t_{1}^{[3]}} \right)  - {\rm ln} \left( 1-
{t_{2}^{[2]}t_{1}^{[4]}\over t_{1}^{[3]}t_{2}^{[3]}} \right) +
\ln\left( - {t_{1}^{[3]}\over t_{2}^{[3]}} \right).}

Even though we have introduced here a special notation for the
monodromy, i.e., $\Delta$, in the rest of the paper we make a
somewhat abuse of notation and call it the imaginary part in the
channel of interest.

\subsec{Four-Mass Scalar Box Function}

As mentioned before, the four-mass scalar box function is special.
It is the only scalar box function that has square roots of the
kinematical invariants in the arguments of the logarithms and
dilogarithms.

Here we consider a given four-mass scalar box function and find to
which cuts it can contribute and the form of the corresponding
monodromy. We then turn to the particular case of the $(1,2,3)$
cut.

Let us consider $F^{4m}_{n:r:r':r'';i}$. The function depends on
six kinematical invariants, i.e. $t_i^{[r]}$,
$t_{i+r+r'}^{[r'']}$, $t_{i+r}^{[r']}$,
$t_{i+r+r'+r''}^{[n-r-r'-r'']}$, $t_{i}^{[r+r']}$, and
$t_{i+r}^{[r'+r'']}$.

Naively, one might think that $F^{4m}_{n:r:r':r'';i}$ has branch
cuts in the six channels defined by the kinematical regime where
anyone of the six invariants is positive and the rest are
negative. This is not possible physically as the box function can
only contribute to four different channels. As we now study in
detail this is indeed the case; in the kinematical regime where
$t_{i+r}^{[r']}$ or $t_{i+r+r'+r''}^{[n-r-r'-r'']}$ are positive
and the rest negative, the four-mass box function does not develop
any imaginary part.

First note that the box function depends on the kinematical
invariants only through the combinations
\eqn\andiw{ \lambda_1={t_i^{[r]}t_{i+r+r'}^{[r"]}\over
t_i^{[r+r']}t_{i+r}^{r'+r"}}, \quad
\lambda_2={t_{i+r}^{[r']}t_{i+r+r'+r"}^{[n-r-r'-r"]}\over
t_i^{[r+r']}t_{i+r}^{r'+r"}}.}

This implies that conditions on the kinematical invariants
translate into conditions on $\lambda_1$ and $\lambda_2$. There
are only three cases to consider:

\item{1.} The $t_i^{[r]}$-cut and the $t_{i+r+r'}^{[r'']}$-cut are
characterized by $\lambda_1 <0$ and $\lambda_2>0$.

\item{2.} The $t_{i+r}^{[r']}$-cut and the
$t_{i+r+r'+r''}^{[n-r-r'-r'']}$-cut are characterized by
$\lambda_1 > 0$ and $\lambda_2 < 0$.

\item{3.} The $t_{i}^{[r+r']}$-cut and the
$t_{i+r}^{[r'+r'']}$-cut are characterized by imposing $\lambda_1
< 0$ and $\lambda_2 < 0$.

We have to consider the arguments of each of the dilogarithms and
logarithms in \woop. For dilogarithms we have to determine whether
the arguments are larger or smaller than one. Recall that
\eqn\kkq{ {\rm Im} ~{\rm Li}_2 (x) = \cases{ 0 & $x\leq 1$ \cr
-\pi \ln(x) & $x>1$ .} }

For the logarithm we have to determine whether the arguments are
positive or negative.

All this discussion has implicitly assumed that the arguments are
real. Note that this is indeed the case. It is not difficult to
check that $\rho$ is always real for any real values of
$\lambda_1$ and $\lambda_2$ in the three regimes of interest.

We find that a table is the most convenient way of presenting all
this information. If a given argument produces an imaginary part
in a given regime we write ``Yes" otherwise we write ``No".
$$\vbox{\settabs
 \+   Box $\lambda_1< 0$ $\lambda_2<0$ &  $-{1\over \lambda_1}K_{--+}$ & ${1\over \lambda_1}K_{--+}$ & $-{1\over \lambda_1}K_{--+}$ &
 $-{1\over \lambda_1}K_{--+}$ & $-{1\over \lambda_1}K_{--+}$ & $-{1\over \lambda_1}K_{--+}$ \cr

\+  &  $K_{-++}~$ & $K_{-+-}~$ & $-{1\over \lambda_1}K_{---}~$ &
 $-{1\over \lambda_1}K_{--+}~$ & $~\lambda_1/ \lambda_2^2$ & $K_{+-+}/ K_{+--}$  \cr

\+ $\lambda_1<0$ $\lambda_2>0$ &  Yes & No & $~~$No &
 $~~$Yes & $~$Yes & Yes   \cr

\+  $\lambda_1> 0$ $\lambda_2<0$  &  No & No & $~~$No &
 $~~$No & $~$No & No \cr

\+   $\lambda_1< 0$ $\lambda_2<0$  &  Yes & No & $~~$No & $~~$Yes
& $~$Yes & No  \cr

 } $$

{}From this table, it is clear that $F^{4m}_{n:r:r':r'';i}$ does
not have an imaginary part neither in the $t_{i+r}^{[r']}$-channel
not in the $t_{i+r+r'+r''}^{[n-r-r'-r'']}$-channel.

Finally, note that for any cut of the form studied in section 2,
we expect the action of the appropriate collinear operator on it
to give a rational function. From our table and \kkq\ it easy to
see that there is no kinematical regime where a pole of the form
$F+\sqrt{K}$ would not be present. This proves that the four-mass
scalar box function can not have a nonzero coefficient if it
participates in this class of cuts.

\appendix{B}{Detailed Computation Of $[F_{123},\eta ] C_{123}$}

In this appendix we provide the details of the computation leading
to \ffool. The computation involves all the details that could be
encountered in the general discussion of section 2 leading to
\excutre. In particular, the jacobian factor ${\cal J}$ is
computed here.

As mentioned in general in section 2 and in our example in section
3, the computation does not involve any actual integration. There
are enough delta functions to localize the integral completely.
Therefore, the only thing to do is to exhibit all delta functions
explicitly, check that their support is in the region of
integration and compute all possible jacobians.

Our starting point here is \kawa,
\eqn\kaap{ \eqalign{ & \int d\mu  {1\over \vev{3~
\ell_2}[\ell_2~4][n~\ell_1]}\vev{2~3}[\del_1,\eta]\left( {1\over
\vev{\ell_1~1}}\right) = \cr & \int d^4\ell_1
\delta^{(+)}(\ell_1^2)\int d^4\ell_2 \delta^{(+)}(\ell_2^2)
\delta^{(4)}(\ell_1+\ell_2-P) {1\over \vev{3~
\ell_2}[\ell_2~4][n~\ell_1]}\vev{2~3}[\del_1,\eta]\left( {1\over
\vev{\ell_1~1}}\right).}   }
Recall that $P=p_1+p_2+p_3$.

The differential operator in this part of the integral only
affects the pole in $\ell_1$. Therefore, it is useful to write the
integral over the future light-cone of $\ell_1$ in terms of spinor
variables. A convenient way of writing this integral\foot{An
alternative possibility can be found in \Nair.} was given in
section 6 of \CachazoKJ. It uses a slightly different
parametrization for null vectors, namely $\ell_1 =
t\lambda_{\ell_1}\tilde\lambda_{\ell_1}$. Then
\eqn\meas{ \int d^4\ell_1 \delta^{(+)}(\ell_1^2) ~ (\bullet ) =
\int_0^{\infty}dt~t\int\vev{\lambda_{\ell_1},
d\lambda_{\ell_1}}[\tilde\lambda_{\ell_1},d\tilde\lambda_{\ell_1}]
( \bullet ),}
where the bullets represent generic arguments. Here
$\lambda_{\ell_1}$ and $\tilde\lambda_{\ell_1}$ are independent
and become homogeneous coordinates for two different
$\Bbb{CP}^1$'s. The integral is a complex integral performed over
the contour $\tilde\lambda_{\ell_1} =\bar\lambda_{\ell_1}$, i.e.,
the diagonal $\Bbb{CP}^1$. On the other and $t$ scales in a way to
make the measure invariant \CachazoKJ.

The different parametrization of $\ell_1$, although convenient, it
requires some care when used. Let us write the original
parametrization as $\ell_1 =
\lambda_{\ell_1}\tilde\lambda_{\ell_1}$. Note that the integrand
of \kaap\ is invariant under simultaneously rescaling
$\lambda_{\ell_1}\to \sigma \lambda_{\ell_1}$ and
$\tilde\lambda_{\ell_1}\to \sigma^{-1} \tilde\lambda_{\ell_1}$.
Therefore, any change of variables from the old to the new
parametrization can be brought to the form $\lambda_{\ell_1, {\rm
old}} =\lambda_{\ell_1,{\rm new}}$ and $\tilde\lambda_{\ell_1,
{\rm old}} =t \tilde\lambda_{\ell_1,{\rm new}}$. This is the
relation we use to convert from one to the other in what follows.

Let us write \kaap\ using the new measure \meas\ and the new
parametrization for $\ell_1$
\eqn\newAll{ \eqalign{ &
\int_0^{\infty}dt~t\int\vev{\lambda_{\ell_1},
d\lambda_{\ell_1}}[\tilde\lambda_{\ell_1},d\tilde\lambda_{\ell_1}]
\int d^4\ell_2\; \delta^{(+)}(\ell_2^2)
\delta^{(4)}(\ell_1+\ell_2-P) \times \cr & {1\over t \vev{3~
\ell_2}[\ell_2~4] [n~\ell_1]}\vev{2~3}[\del_1,\eta]\left( {1\over
\vev{\ell_1~1}}\right).}   }
Note that a factor of $t$ has appeared in the denominator.

Having written the integral in a convenient form, let us turn to
the holomorphic anomaly. Using \really\ with $\lambda =
\lambda_{\ell_1}$ and $\lambda' = \lambda_1$, we have
\eqn\real{ d\bar\lambda_{\ell_1}^{\dot
a}{\partial\over\partial\bar\lambda_{\ell_1}^{\dot a}}{1\over
\langle\lambda_{\ell_1},\lambda_1\rangle}= [d\bar\lambda_{\ell_1},
\del_{\ell_1}]{1\over \langle\lambda_{\ell_1},\lambda_1\rangle}
=2\pi\bar\delta(\langle\lambda_{\ell_1},\lambda_1\rangle). }
This form is useful because the delta function can be used to
carry out the integral over $\lambda_{\ell_1}$ and
$\tilde\lambda_{\ell_1}$ in \newAll.

However, the operator acting on the pole in \newAll\ is not of the
form required in \real. To bring the operator to that form note
the following identity
\eqn\uno{  [\del_1,\eta]\left( {1\over \vev{\ell_1~1}}\right) = -
[\del_{\ell_1},\eta]\left( {1\over \vev{\ell_1~1}}\right) . }
Inserting this in \newAll\ produces a factor of the form
$[\tilde\lambda_{\ell_1},
d\tilde\lambda_{\ell_1}][\del_{\ell_1},\eta]$ which by Schouten's
identity \schouten\ becomes
\eqn\soto{ [\tilde\lambda_{\ell_1}, d\tilde\lambda_{\ell_1}]
[\del_{\ell_1},\eta] =
[\tilde\lambda_{\ell_1},\del_{\ell_1}][d\tilde\lambda_{\ell_1},\eta]
- [\tilde\lambda_{\ell_1},\eta
][d\tilde\lambda_{\ell_1},\del_{\ell_1}] }
Note that the second term in \soto\ is precisely what we want in
order to use \real. Luckily, the first term does not contribute.
To see this note that in the contour of integration
\eqn\woops{[\bar\lambda_{\ell_1},\del_{\ell_1}]\left( {1\over
\vev{\ell_1~1}}\right) =
[\bar\lambda_{\ell_1},\bar\lambda_1]\delta(\langle\lambda_{\ell_1},\lambda_1\rangle)
=[\bar\lambda_1,\bar\lambda_1]\delta(\langle\lambda_{\ell_1},\lambda_1\rangle)=0.
}

Let us collect all our partial result to write \newAll\ as follows
\eqn\kafo{\int d^4\ell_2 \delta^{(+)}(\ell_2^2) \int_0^{\infty}dt~
\int
\vev{\lambda_{\ell_1},d\lambda_{\ell_1}}\bar\delta(\langle\lambda_{\ell_1},\lambda_1\rangle)
\delta^{(4)}(\ell_1+\ell_2-P){\vev{2~3}[\bar\lambda_{\ell_1},\eta]\over
\vev{3~\ell_2}[\ell_2~4][n~\ell_1]} }
Now the integral over $\lambda_{\ell_1}$ can be evaluated
trivially by using the delta function to set
$\lambda_{\ell_1}=\lambda_1$ and therefore $\ell_1 = t p_1$. After
this is done we are left with
\eqn\wuwu{ \int d^4\ell_2 \delta^{(+)}(\ell_2^2) \int_0^{\infty}dt
\;\delta^{(4)}(t p_1+\ell_2-P){\vev{2~3}[1,\eta]\over
\vev{3~\ell_2}[\ell_2~4][n~1]}.}

The integral over $\ell_2$ is trivial because we can use
$\delta^{(4)}(t p_1 + \ell_2-P)$ to get
\eqn\alti{{\vev{2~3}[1~\eta]\over [n~1]}\int_0^{\infty}dt\;
\delta^{(+)}(\ell_2^2) \;{1 \over \vev{3~\ell_2}[\ell_2 ~ 4]} }
where $\ell_2 = P - tp_1$.

The remaining delta function can be written in a more convenient
form
\eqn\remax{ \delta^{(+)}(\ell_2^2) = \delta \left( t (2 p_1\cdot
P)-P^2 \right) \vartheta (E_{\ell_2})}
where $E_{\ell_2}$ is the energy component of $\ell_2$ and
$\vartheta (x)$ is $1$ for $x\ge 0$ and $0$ for $x<0$.

Note that $(2 p_1\cdot P) = t_1^{[3]} - t_2^{[2]}$. Recalling that
this computation is done in the kinematical regime where
$t_1^{[3]}>0$ and all other invariants are negative, in particular
$t_2^{[2]}<0$, it is clear that $(2 p_1\cdot P)$ is always
positive. Therefore it can easily be pulled out of the delta
function
\eqn\formal{ \delta^{(+)}(\ell_2^2) = {1\over 2 p_1\cdot
P}\delta\left( t - {P^2\over 2 p_1\cdot P} \right)\vartheta
(E_{\ell_2}).}
The integral over $t$ in \alti\ is again trivial and nonzero, for
$t = t_1^{[3]}/(t_1^{[3]}-t_2^{[2]})$ always satisfies the
following condition: $0< t < 1$. This also implies that
$E_{\ell_2}$ is always positive. To see this recall that $\ell_2 =
(1-t)p_1+p_2+p_3$ and that all $p_1$, $p_2$, and $p_3$ have
positive energy.

Combining all this we find that \alti\ equals
\eqn\jodu{{\vev{2~3}[1~\eta]\over [n~1]\vev{3~\ell_2}[\ell_2 ~
4](2 p_1\cdot P)}. }
Writing $[\ell_2 ~ 4] = (2p_4\cdot \ell_2)/\vev{\ell_2~4}$ we get
\eqn\joda{\vev{2~3}{[1~\eta]\over [n~1]}{\vev{4~\ell_2}\over
\vev{3~\ell_2}}\; {1\over  (2p_1\cdot P)(2p_4\cdot P)-P^2
(2p_1\cdot p_4)}. }
This formula can be further simplified by noticing the following
identities
\eqn\tyty{ (2p\cdot P)(2q\cdot P) - P^2 (2p\cdot q) = {\langle q
|P|p]}{\langle p|P|q ]}}
and
\eqn\kuku{ {\vev{4~\ell_2}\over \vev{3~\ell_2}} = {{\langle 4
|P|1]} \over {\langle 3 |P|1]}} ={{\langle 4 |P|1]} \over
\vev{3~2}[2~1]} .}
Using \tyty\ with $p=p_4$ and $q=p_1$, and \kuku\ in \joda\ we
finally find \ffool\
\eqn\ffoo{\int d\mu  {1\over \vev{3~
\ell_2}[\ell_2~4][n~\ell_1]}\vev{2~3}[\del_1,\eta]\left( {1\over
\vev{\ell_1~1}}\right)  =  {[1~\eta]\over {\langle 1
|P|4]}[n~1][1~2]}. }

\listrefs

\end